\documentclass{jfm}
\usepackage{graphicx}
\usepackage{epstopdf, epsfig}
\usepackage{subfigure}
\usepackage{amsmath}
\usepackage{amstext}

\shorttitle{Sub-grid modelling using neural networks}
\shortauthor{R. Maulik, O. San, A. Rasheed, and P. Vedula}

\title{Sub-grid modelling for two-dimensional turbulence using neural networks}

\author{R. Maulik\aff{1},
  O. San\aff{1}\corresp{\email{osan@okstate.edu}}, A. Rasheed\aff{2}, P. Vedula\aff{3} }

\affiliation{
\aff{1}School of Mechanical \& Aerospace Engineering, Oklahoma State University, Stillwater, OK 74078, USA
\aff{2}CSE Group, Applied Mathematics and Cybernetics, SINTEF Digital, N-7465 Trondheim, Norway
\aff{3}School of Aerospace \& Mechanical Engineering, The University of Oklahoma Norman, OK 73019, USA
}

\begin{document}

\maketitle

\begin{abstract}
In this investigation, a data-driven turbulence closure framework is introduced and deployed for the sub-grid modelling of Kraichnan turbulence. The novelty of the proposed method lies in the fact that snapshots from high-fidelity numerical data are used to inform artificial neural networks for predicting the turbulence source term through localized grid-resolved information. In particular, our proposed methodology successfully establishes a map between inputs given by stencils of the vorticity and the streamfunction along with information from two well-known eddy-viscosity kernels. Through this we predict the sub-grid vorticity forcing in a temporally and spatially dynamic fashion. Our study is both a-priori and a-posteriori in nature. In the former, we present an extensive hyper-parameter optimization analysis in addition to learning quantification through probability density function based validation of sub-grid predictions. In the latter, we analyse the performance of our framework for flow evolution in a classical decaying two-dimensional turbulence test case in the presence of errors related to temporal and spatial discretization. Statistical assessments in the form of angle-averaged kinetic energy spectra demonstrate the promise of the proposed methodology for sub-grid quantity inference. In addition, it is also observed that some measure of a-posteriori error must be considered during optimal model selection for greater accuracy. The results in this article thus represent a promising development in the formalization of a framework for generation of heuristic-free turbulence closures from data.
\end{abstract}


\section{Introduction}

The efficient computational modelling of energetic flows continues to remain an important area of research for many engineering and geophysical applications. Over the past few decades, coarse-grained techniques such as Reynolds-averaged Navier-Stokes (RANS) and large eddy simulation (LES) have proven promising for the statistically accurate prediction of the grid-resolved scales of a turbulent flow. While RANS is based on the modelling of turbulence in a temporally averaged sense, LES requires the specification of a model for the finer scales and their effect on the grid-resolved quantities. This modelling of the excluded wavenumbers in LES represents the classical closure problem which has spawned a variety of algebraic or equation based techniques for representing the effect of these discarded scales on the resolved ones \citep{berselli2005mathematics,sagaut2006large}. It has generally been observed that the choice of the sub-grid model is physics dependant, i.e., that different flow phenomena require different expressions for sub-grid terms with a-priori assumptions of phenomenology \citep{vreman2004eddy}. We use this fact as a motivation for moving to an equation-free model for the source term through the use of an artificial neural network (ANN). Our hope, in addition to the formulation of a prediction framework, is to devise the formalization of a `machine-learning experiment' where a-priori model selection and a-posteriori deployment are coupled to reveal information about the physical characteristics of a particular flow class. This not only enables the selection of computationally efficient predictive models but also reveals the importance of certain grid-resolved quantities of interest from the flow characteristics. In accordance with the recent trends of first-principles informed learning for physics inference in turbulence \citep{ling2015evaluation,tracey2015machine,xiao2016quantifying,singh2017machine,wang2017physics,wang2017comprehensive,weatheritt2017hybrid,schaeffer2017learning,wu2018data,raissi2018hidden,wan2018data,mohan2018deep}, a major goal of this research is to study the combination of the traditional learning framework (inherently data-driven) and the physics-based prediction tool (based on the coarse-grained Navier-Stokes equations). We devote particular attention to the necessity for physical realizability as well as the issues faced by learning frameworks and their interactions with numerical discretization error.

Over the past decade, there have been multiple studies on the use of machine learning tools for the reduced-order prediction of energetic flow physics. The study of these techniques has been equally popular for both severely truncated systems such as those obtained by leveraging sparsity in transformed bases \citep{faller1997unsteady,cohen2003feedback,mannarino2014nonlinear,san2018neural} as well as for modelling methodologies for coarse-grained meshes such as LES and RANS simulations \citep{maulik2017neural,wang2017physics,wu2018physics}. Therefore they represent a promising direction for the assimilation of high-fidelity numerical and experimental data during the model-formulation phase for improved predictions during deployment. A hybrid formulation leveraging our knowledge of governing equations and augmenting these with machine learning represents a great opportunity for obtaining optimal LES closures for multiscale physics simulations \citep{langford1999optimal,moser2009theoretically,king2016autonomic,pathak2018hybrid}.

From the point of view of turbulence modelling, we follow a strategy of utilizing machine learning methods for estimating the sub-grid forcing quantity such as the one utilized in \cite{ling2016reynolds} where a deep ANN has been described for Reynolds stress predictions in an invariant subspace. ANNs have been also implemented in \cite{parish2016paradigm} to correct errors in RANS turbulence models after the formulation of a field-inversion step. \cite{gamahara2017searching} detailed the application of ANNs for identifying quantities of interest for sub-grid modelling in a turbulent channel flow through the measurement of Pearson correlation coefficients. \cite{milano2002neural} also implemented these techniques for turbulent channel flow but for the generation of low-order wall models while \cite{sarghini2003neural} deployed ANNs for the prediction of the Smagorinsky coefficient (and thus the sub-grid contribution) in a mixed sub-grid model. In \cite{beck2018neural}, an ANN prediction has been hybridized with a least-squares projection onto a truncated eddy-viscosity model for LES. In these (and most) utilizations of machine learning techniques, sub-grid effects were estimated using grid-resolved quantities. Our approach is similar, wherein grid-resolved information is embedded into the input variables for predicting LES source terms for the filtered vorticity transport equation.

We outline a methodology for the development, testing and validation of a purely data-driven LES modelling strategy using ANNs which precludes the utilization of any phenomenology. However, in our framework the machine learning paradigm is used for predicting the vorticity forcing or damping of the unresolved scales, which lends to an easier characterization of numerical stability restrictions as well as ease of implementation. Our model development and testing framework is outlined for Kraichnan turbulence \citep{kraichnan1967inertial} where it is observed that a combination of a-priori and a-posteriori analyses ensure the choice of model frameworks that are optimally accurate and physically constrained during prediction. Conclusions are drawn by statistical comparison of predictions with high-fidelity data drawn from direct numerical simulations (DNS).

To improve the viability of our proposed ideas, we devise our learning using extremely sub-sampled data sets. The use of such sub-sampled data necessitates a greater emphasis on physics-distillation to prevent extrapolation and over-fitting during the training phase. An a-priori hyper-parameter optimization is detailed for the selection of our framework architecture before deployment. An a-posteriori prediction in a numerically evolving flow tests the aforementioned `learning' of the framework for spectral scaling recovery which are compared to robust models utilizing algebraic eddy-viscosities given by the Smagorinsky \citep{smagorinsky1963general} and Leith \citep{leith1968diffusion} models. A hardwired numerical realizability also ensures viscous stability of the proposed framework in an a-posteriori setting. Later discussions demonstrate how the proposed framework is suitable for the prediction of vorticity forcing as well as damping in the modeled scales. The proposed formulation also ensures data-locality, where a dynamic forcing or dissipation of vorticity is specified spatio-temporally.

Following our primary assessments, our article proposes the use of a combined a-priori and a-posteriori study for optimal predictions of kinetic energy spectra as well as hyper-parameter selection prior to deployment for different flows which belong to the same class but have a different control parameter or initial conditions. It is also observed that the specification of eddy-viscosity kernels (which are devised from dimensional analyses) constrain the predictive performance of the framework for the larger scales. Results also detail the effect of data-locality, where an appropriate region of influence utilized for sampling is shown to generate improved accuracy. The reader may find a thorough review of concurrent ideas in \cite{duraisamy2018turbulence}. An excellent review of the strengths and opportunities of using artificial neural networks for fluid dynamics applications may also be found in \cite{kutz2017deep}.

The mathematical background of sub-grid modelling for the LES of two-dimensional turbulence may be summarized in the following. In terms of the vorticity-streamfunction formulation, our non-dimensional governing equation for incompressible flow may be represented as
\begin{align}
\label{eq1}
\frac{\partial \omega}{\partial t} + J(\omega,\psi) = \frac{1}{Re} \nabla^2 \omega,
\end{align}
where $Re$ is the Reynolds number, $\omega$ and $\psi$ are the vorticity and streamfunction respectively connected to each other through the Poisson equation given by
\begin{align}
\label{eq2}
\nabla^2 \psi = - \omega.
\end{align}
It may be noted that the Poisson equation implicitly ensures a divergence-free flow evolution. The nonlinear term (denoted the Jacobian) is given by
\begin{align}
\label{eq3}
J(\omega,\psi) = \frac{\partial \psi}{\partial y} \frac{\partial \omega}{\partial x} - \frac{\partial \psi}{\partial x} \frac{\partial \omega}{\partial y}.
\end{align}

A reduced-order implementation of the aforementioned governing laws (i.e., an LES) is obtained through
\begin{align}
\label{eq4}
\frac{\partial \bar{\omega}}{\partial t} + J(\bar{\omega},\bar{\psi}) = \frac{1}{Re} \nabla^2 \bar{\omega} + \Pi,
\end{align}
where the overbarred variables are now evolved on a grid with far fewer degrees of freedom. The sub-grid term $\Pi$ encapsulates the effects of the finer wavenumbers which have been truncated due to insufficient-grid support and must be approximated by a model. Mathematically we may express this (ideal) loss as
\begin{align}
\label{eq5}
\Pi = J(\bar{\omega},\bar{\psi}) - \overline{J(\omega,\psi)}.
\end{align}
In essence, the basic principle of LES is to compute the largest scales of turbulent motion and use closures to model the contributions from the smallest turbulent flow scales. The nonlinear evolution equations introduce unclosed terms that must be modeled to account for local, instantaneous momentum and energy exchange between resolved and unresolved scales. If these inter-eddy interactions are not properly parameterized, then an increase in resolution will not necessarily improve the accuracy of these large scales \citep{frederiksen2016theoretical,frederiksen2013subgrid}. Additionally, most LES closures are based on three-dimensional turbulence considerations primarily encountered in engineering applications. These LES models fundamentally rely on the concept of the forward energy cascade and their extension to geophysical flows is challenging \citep{eden2008towards,fox2011parameterization,san2013approximate}, due to the effects of stratification and rotation which suppress vertical motions in the thin layers of fluid. In the following, we shall elaborate on the use of a machine learning framework to predict the approximate value of $\Pi$ in a pointwise fashion on the coarser grid and assess the results of its deployment in both a-priori and a-posteriori testing. Through this we attempt to bypass an algebraic or differential equation based specification of the turbulence closure and let the data drive the quantity and quality of sub-grid forcing. We note here that the definition of the sub-grid source term given in Equation \ref{eq5} is formulated for the LES of two-dimensional Navier-Stokes equations in the vorticity-streamfunction formulation but the framework outlined in this article may be readily extended to the primitive-variable formulation in two or higher dimensions \citep{mansfield1998dynamic,marshall2003analysis}.

\section{Machine learning architecture}

\subsection{Mathematical formulation}

\begin{figure}
\centering
\includegraphics[width=0.90\textwidth]{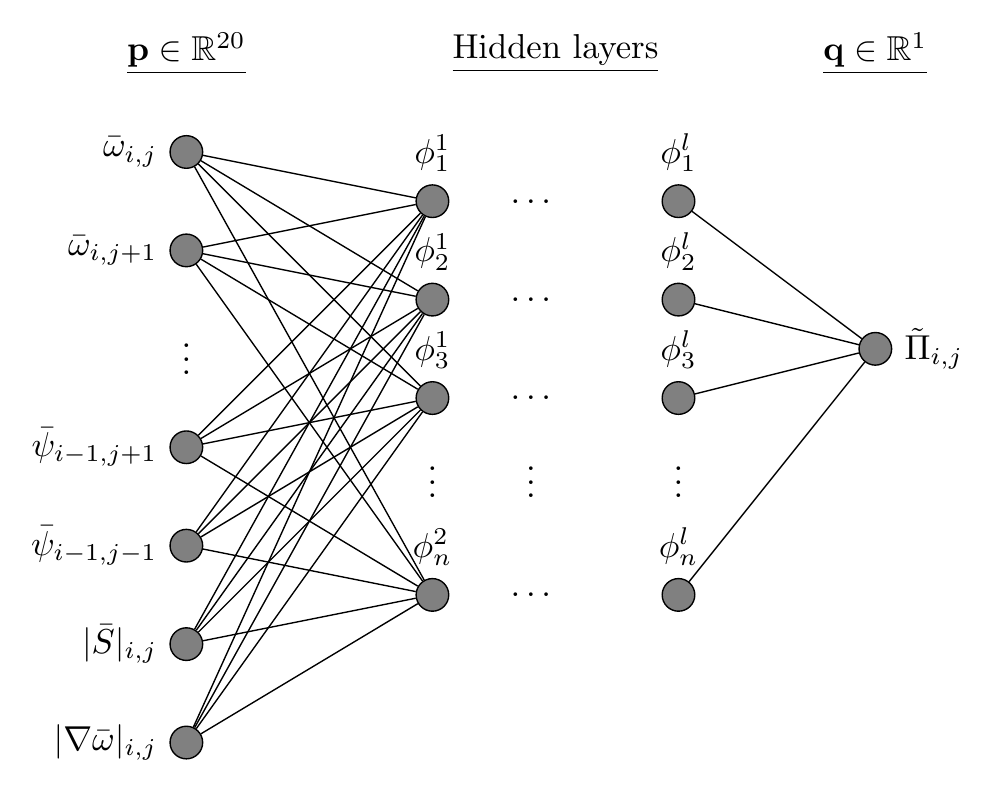}
\caption{Proposed artificial neural network architecture and relation to sampling and prediction space.}
\label{fig:fig1}
\end{figure}

In this section, we introduce the machine learning methodology employed for the previously described regression problem. The ANN, also known as a multilayered perceptron, consists of a set of linear or nonlinear mathematical operations on an input space vector to establish a map to an output space. Other than the input and output spaces, an ANN is also said to contain multiple hidden layers (denoted so due to the obscure mathematical significance of the matrix operations occurring here). Each of these layers is an intermediate vector in a multi-step transformation which is acted on by biasing and activation before the next set of matrix operations. Biasing refers to an addition of a constant vector to the incident vector at each layer, on its way to a transformed output. The process of activation refers to an element-wise functional modification of the incident vector to generally introduce nonlinearity into the eventual map. In contrast, no activation (also referred to as `linear' activation), results in the incident vector being acted on solely by biasing. Note that each component of an intermediate vector corresponds to a unit cell also known as the neuron. The learning in this investigation is \emph{supervised} implying label data used for informing the optimal map between inputs and outputs. Mathematically, if our input vector $\textbf{p}$ resides in a $P$-dimensional space and our desired output $\textbf{q}$ resides in a $Q$-dimensional space, this framework establishes a map $\mathbb{M}$ as follows:
\begin{align}
\label{eq6}
\mathbb{M} : \{ p_1, p_2, \hdots, p_P\} \in \mathbb{R}^P \rightarrow \{ q_1, q_2, \hdots, q_Q\} \in \mathbb{R}^Q.
\end{align}
A schematic for this map may be observed in Figure \ref{fig:fig1}, where input, output and hidden spaces are summarized. In equation form, our default optimal map is given by
\begin{align}
\label{eq7}
\begin{gathered}
\mathbb{M} : \{ \bar{\omega}_{i,j}, \bar{\omega}_{i,j+1}, \bar{\omega}_{i,j-1}, \hdots, \bar{\omega}_{i-1,j-1}, \\ \bar{\psi}_{i,j}, \bar{\psi}_{i,j+1}, \bar{\psi}_{i,j-1}, \hdots, \bar{\psi}_{i-1,j-1}, |\bar{S}|_{i,j}, |\nabla \bar{\omega}|_{i,j}\} \in \mathbb{R}^{20} \rightarrow \{ \tilde{\Pi}_{i,j}\} \in \mathbb{R}^1.
\end{gathered}
\end{align}
where
\begin{align}
\label{eq8}
|\bar{S}| = \sqrt{\strut 4 \left( \frac{\partial^2 \bar{\psi}}{\partial x \partial y}\right)^2 + \left( \frac{\partial^2 \bar{\psi}}{\partial x^2} - \frac{\partial^2 \bar{\psi}}{\partial y^2}\right)^2}, \quad
|\nabla \bar{\omega}| = \sqrt{\strut \left(\frac{\partial \bar{\omega}}{\partial x}\right)^2 + \left(\frac{\partial \bar{\omega}}{\partial y}\right)^2}
\end{align}
are eddy-viscosity kernel information input to the framework and $\tilde{\Pi}$ is the approximation to the true sub-grid source term. Note that the indices $i$ and $j$ correspond to discrete spatial locations on a coarse-grained two-dimensional grid. The map represented by Equation \ref{eq7} is considered `default' due to the utilization of a 9-point sampling stencil of vorticity and streamfunction (corresponding to 18 total inputs) and two other inputs of the Smagorinsky and Leith kernels. The purpose of utilizing the additional information from these well-established eddy-viscosity hypotheses may be considered a data pre-processing mechanism where certain important quantities of interest are distilled and presented `as-is' to the network for simplified architectures and reduced training durations. The motivation behind the choice of these particular kernels is discussed in later sections where it is revealed that they also introduce a certain regularization to the optimization. We note that all our variables in this study are non-dimensionalized at the stage of problem definition and no further pre-processing is utilized prior to exposing the map to the input data for predictions. The predicted value of $\tilde{\Pi}$ is post-processed before injection into the vorticity equation as follows:
\begin{align}
\label{eq9}
\Pi =
\begin{cases}
\tilde{\Pi},& \text{if  } (\nabla^2 \bar{\omega}) (\tilde{\Pi}) > 0\\
    0,              & \text{otherwise.}
\end{cases}
\end{align}
This ensures numerical stability due to potentially negative eddy-viscosities embedded in the source term prediction and may be considered to be an implicit assumption of Bousinessq hypothesis for functional sub-grid modelling. It is later demonstrated that the presence of this constraint does not preclude the prediction of positive or negative values of $\tilde{\Pi}$, which implies that the proposed framework is adept at predicting vorticity forcing or damping at the finer scales respectively. The damping of vorticity at the finer scales would correspond to a lower dissipation of kinetic energy (assuming that vorticity dissipates kinetic energy in the sub-grid scales). Similarly, the forcing of vorticity at the finer scales may be assumed to be an localized event of high kinetic energy dissipation. In general, Equation \ref{eq9} precludes the presence of a backscatter of enstrophy for strict adherence to viscous stability requirements on the coarse-grained mesh. Instead of the proposed truncation, one may also resort to some form of spatial averaging in an identifiable homogeneous direction as utilized by \cite{germano1991dynamic}. However, the former was chosen to remove any dependency on model-forms or coefficient calculations. In what follows for the rest of this document, our proposed framework is denoted ANN-SGS.

\subsection{Hyper-parameter optimization}

In this sub-section, we detail the process of a-priori architecture selection before training and deployment. Our hidden layers have neurons which are activated by the rectified-linear (ReLU) function. The choice of the ReLU activation was made for efficient optimization of the network architecture by bypassing the problems of vanishing gradients inherent in sigmoidal activation functions \citep{ling2016reynolds}.

\begin{figure}
\centering
\mbox{
\subfigure{\includegraphics[width=0.9\textwidth]{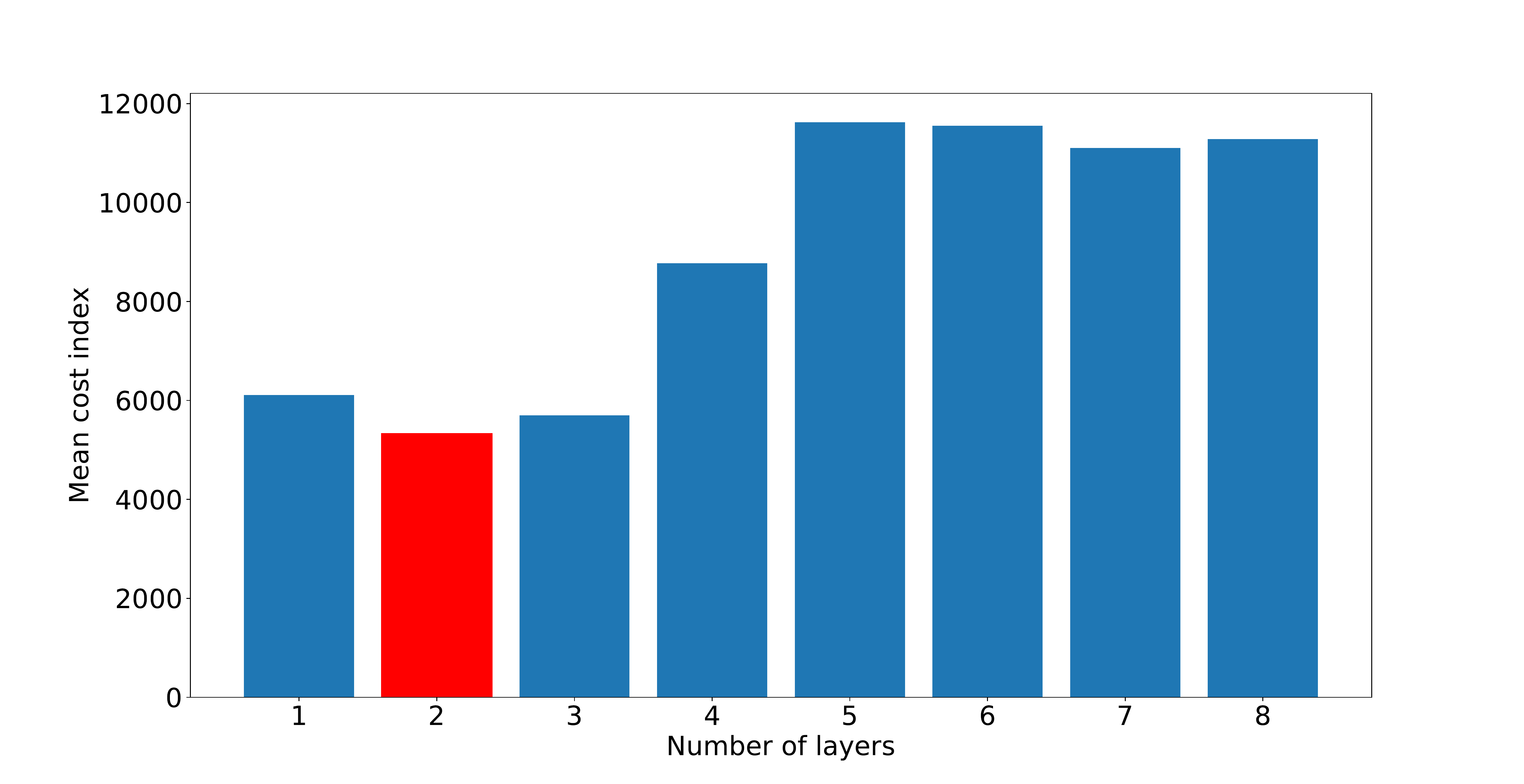}}
} \\
\mbox{
\subfigure{\includegraphics[width=0.9\textwidth]{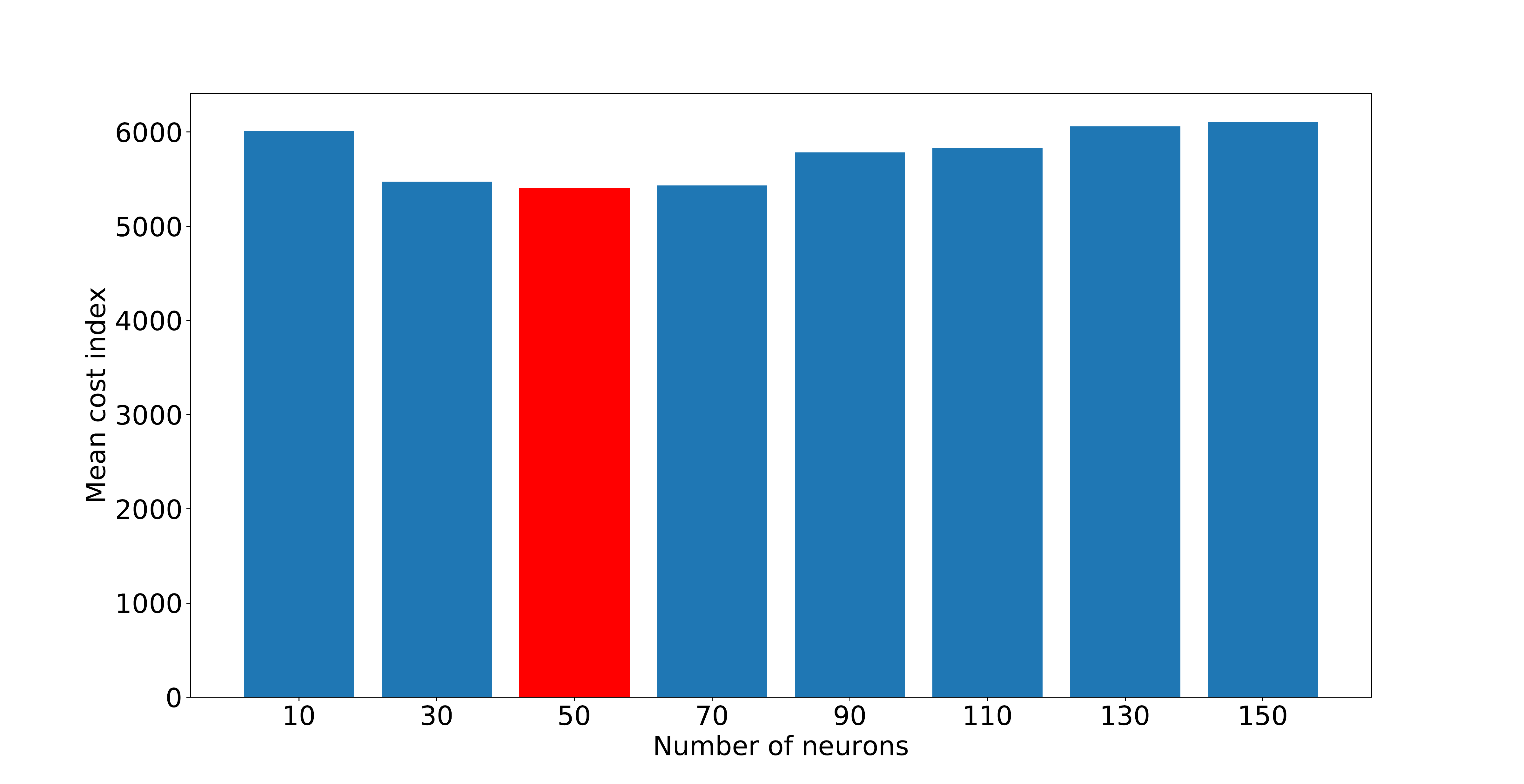}}
}
\caption{Quantification of hyper-parameter optimization shown for number of layers (top) and number of neurons (bottom). An optimal network architecture of two-layers and 50 neurons is chosen for our study.}
\label{fig:fig2}
\end{figure}

For the purpose of optimal network architecture selection, we utilize a grid-search selection coupled with a 3-fold cross-validation implemented in the open-source library Scikit-learn. In essence, a parameter space given by a grid is coupled with three trainings, tests and validations for each network through three partitions of the total training data. We first undertake our aforementioned optimization for the number of layers by utilizing a total of 1000 epochs for determining the optimal depth of the network. Each network with a particular choice of the number of layers (ranging between 1 to 8) is optimized three times using a 3-fold cross-validation strategy and utilized for prediction on the test and validation partitions not used for weight optimization. The three networks for each hyper-parameter are then assigned a mean cost-function score which is used for selection of the final model depth. We observe that a two-layer model outperforms other alternatives during this grid-search as shown in Figure \ref{fig:fig2}. We note that the number of neurons in this first grid-search is fixed at 50 although similar trends are recovered with varying specifications between 10 and a 100. Our mean cost index is given by the following expression for each location on the grid
\begin{align}
\text{Mean cost index} = \frac{1}{K} \sum_{i=1}^K \left| \left| \Pi^{true}_K - \tilde{\Pi}_K \right|\right|_2
\end{align}
where $K$ refers to the training fold chosen for gradient calculation in the backpropagation within the same dataset.

\begin{figure}
\centering
\includegraphics[width=0.90\textwidth]{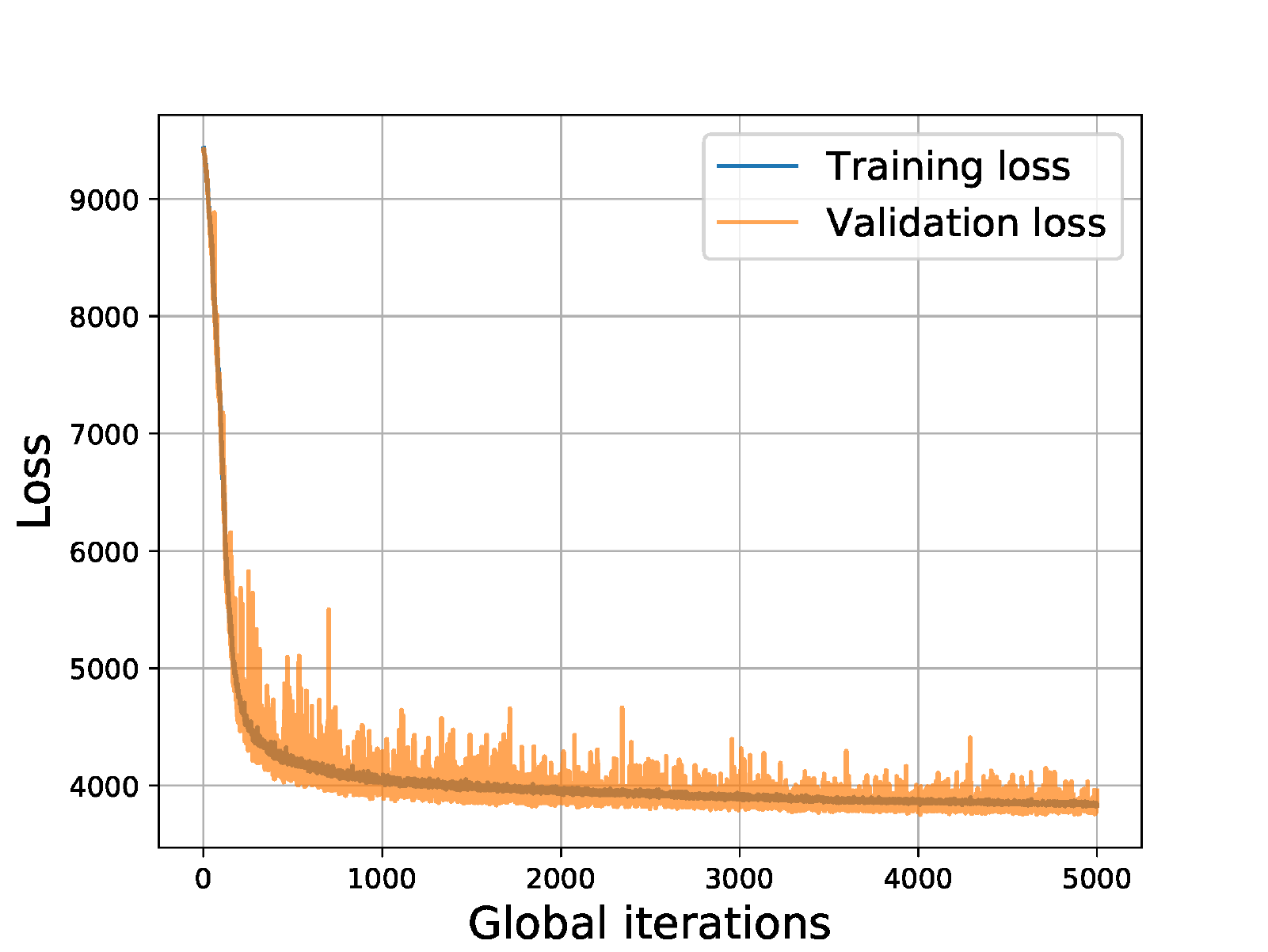}
\caption{Learning rate of the proposed optimal model architecture. Note how training and validation loss are correlated closely for this learning problem.}
\label{fig:fig3}
\end{figure}

A second grid-search is performed with a fixed number of layers (i.e., two obtained from the previous tuning) and with a varying number of neurons. The results of this optimization are observed in Figure \ref{fig:fig2} which shows that an optimal number of neurons of 50 suffice for this training. We note however, that the choice for the number of neurons in the two-layer network does not affect the tuning score significantly. We clarify here that the model optimization may have been carried out using a multidimensional grid-search for the optimal hyper-parameters or through sampling in a certain probability distribution space, however our approach was formulated out of a desire to reduce offline training cost as much as possible. The final network was then selected for a longer duration of training (5000 epochs) till the learning rate is minimal as shown in Figure \ref{fig:fig3}. Details of our network optimization and dataset generation are provided in the next section.

\section{Training and validation}

For the purpose of generating an optimal map discussed in the previous section, we utilize a supervised learning with sets of labeled inputs and outputs obtained from direct numerical simulation data (DNS) for two-dimensional turbulence \citep{san2012high,maulik2017stable}. Our grid-resolved variables (which we remind the reader, are denoted as overbarred quantities) are generated by a Fourier cut-off filter so as to truncate the fully-resolved DNS fields (obtained at $2048^2$ degrees-of-freedom) to coarse-grained grid level (i.e. given by $256^2$ degrees-of-freedom). Therefore, this procedure is utilized to generate input-output pairs for the process of training our ANN map. We also emphasize on the fact that, while the DNS data generated multiple time snapshots of flow evolution, data was harvested from times $t=0,1,2,3$ and $4$ for the purpose of training and validation. This represents a stringent sub-sampling of the total available data for map optimization. To quantify this sub-sampling, we note that we had potential access to 40000 space-time snapshots of DNS data out of which only 5 were chosen for training and validation data generation (0.0125 \% of total data). We also note that the Reynolds number chosen for generating the training and validation data sets is given by $Re=32000$ alone.

Two-thirds of the total dataset generated for optimization was utilized for training and the rest was utilized for validation assessment. Here, training refers to the use of data for loss calculation (which in this study is a classical mean-squared-error) and backpropagation for parameter update. Validation was utilized to record the performance of the trained network on data it was not exposed to during training. Similar behavior in training and validation loss would imply a well-formulated learning problem. The final ANN (obtained post-training) would be selected according to the best validation loss after a desired number of iterations which for this study was fixed at 5000. We also note that the error-minimization in the training of the ANN utilized the Adam optimizer \citep{kingma2014adam} implemented in the open-source ANN training platform TensorFlow. Figure \ref{fig:fig3} shows the learning rate of the proposed framework with very similar behavior between training and validation loss implying a successfully optimized map. We remark that while the network may have learned the map from the data it has been provided for training and validation, testing would require an a-posteriori examination as detailed in the following section.

We first outline an a-priori study for the proposed framework where the optimal map is utilized for predicting probability distributions for the true sub-grid source term. In other words, we assess the turbulence model for a one snapshot prediction. Before proceeding, we return to our previous discussion about the choice of Smagorinsky and Leith viscosity kernels by highlighting their behavior for different choices of model coefficients (utilized in effective eddy-viscosity computations using mixing-length based phenomenological arguments). The Smagorinsky or Leith sub-grid scale models may be implemented in the vorticity-streamfunction formulation via the specification of an effective eddy-viscosity
\begin{align}
\label{eq10}
	\tilde{\Pi} = \nu_e \nabla^2 \bar{\omega},
\end{align}
where the Smagorinsky model utilizes
\begin{align}
\label{11}
\nu_e = (C_s \delta)^2 |\bar{S}|,
\end{align}
while the Leith hypothesis states
\begin{align}
\label{eq12}
	\nu_e = (C_l \delta)^3 |\nabla \bar{\omega}|.
\end{align}
In the above relations, $\delta$ refers to the grid-volume (or area in two-dimensional cases) and $\nu_e$ is an effective eddy-viscosity. From Figure \ref{fig:fig1}, it is apparent that the choice of model-form coefficients $C_s$ and $C_l$ for the Smagorinsky and Leith models dictate the accuracy of the closure model in a-priori analyses. Instances here refer to the probability densities of truth and prediction at different magnitudes. We would also like to draw the readers attention to the fact that ideal reconstructions of the true sub-grid term are with coefficients near the value of 1.0, a value that is rather different to the theoretically accepted values of $C_s$ applicable in three-dimensional turbulence. This dependance of closure efficacy on model coefficients continues to represent a non-trivial a-priori parameter specification task for practical utilization of common LES turbulence models particularly in geophysical applications. Later, we shall demonstrate that a-posteriori implementations of these static turbulence models is beset with difficulties for non-stationary turbulent behavior.

\begin{figure}
\centering
\mbox{
\subfigure[Smagorinsky]{\includegraphics[width=0.44\textwidth]{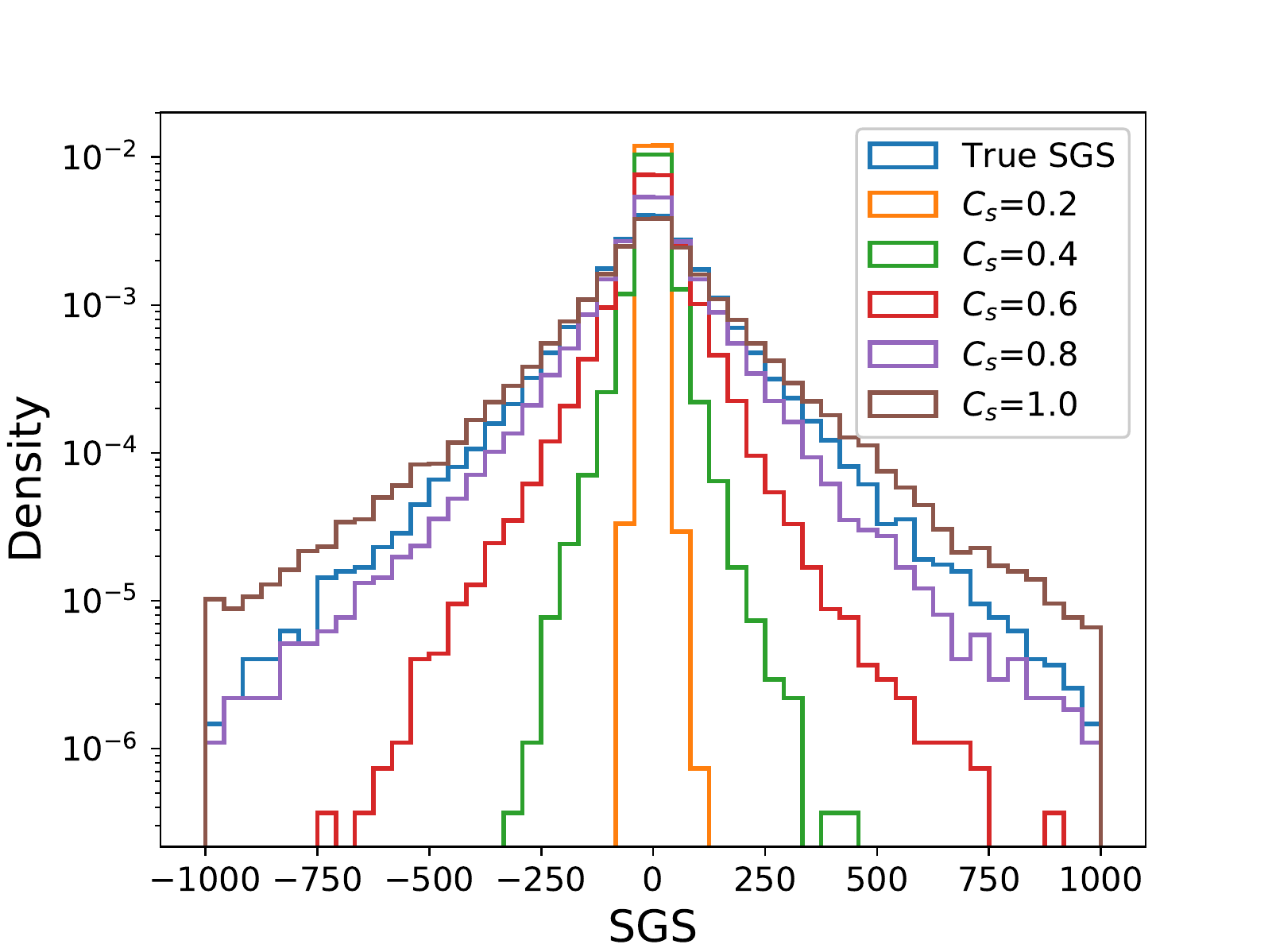}}
\subfigure[Leith]{\includegraphics[width=0.44\textwidth]{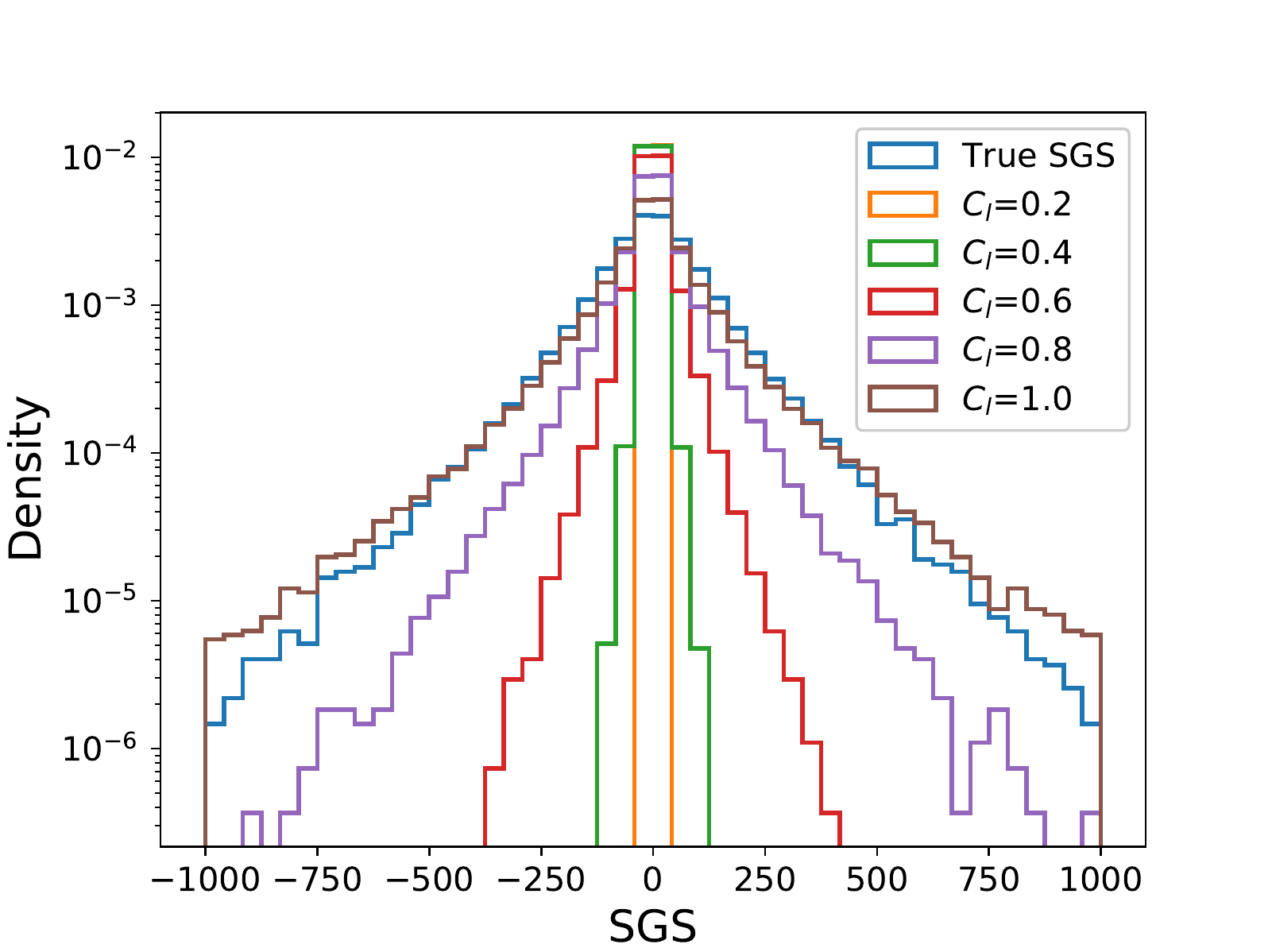}}
}
\caption{A-priori performance of Smagorinsky and Leith models for varying model coefficients for data snapshot at $t=2$. Here, instances refer to the probability densities of truth and prediction at different magnitudes.}
\label{fig:fig4}
\end{figure}

\begin{figure}
\centering
\mbox{
\subfigure[$Re=32000$]{\includegraphics[width=0.44\textwidth]{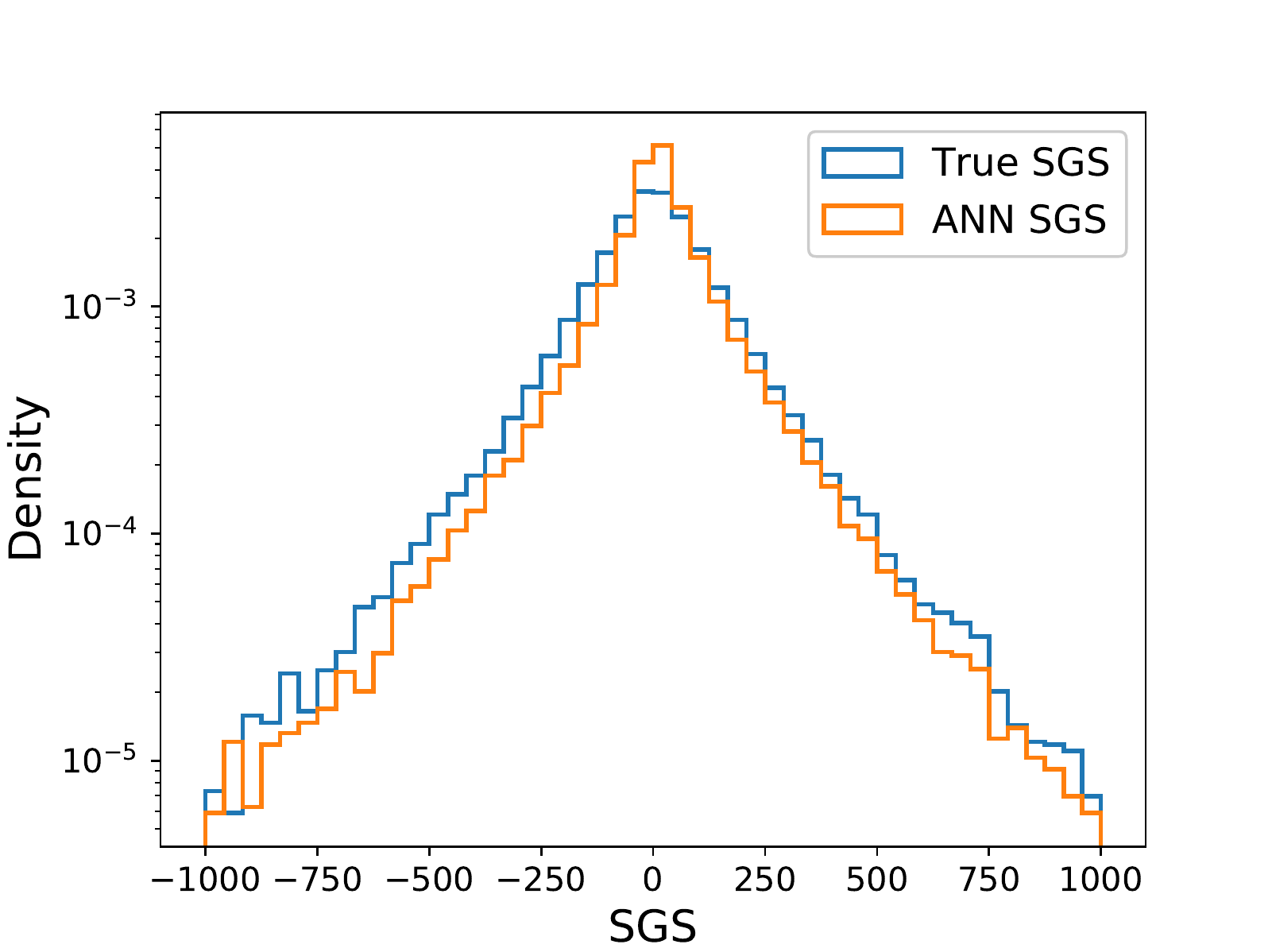}}
\subfigure[$Re=64000$]{\includegraphics[width=0.44\textwidth]{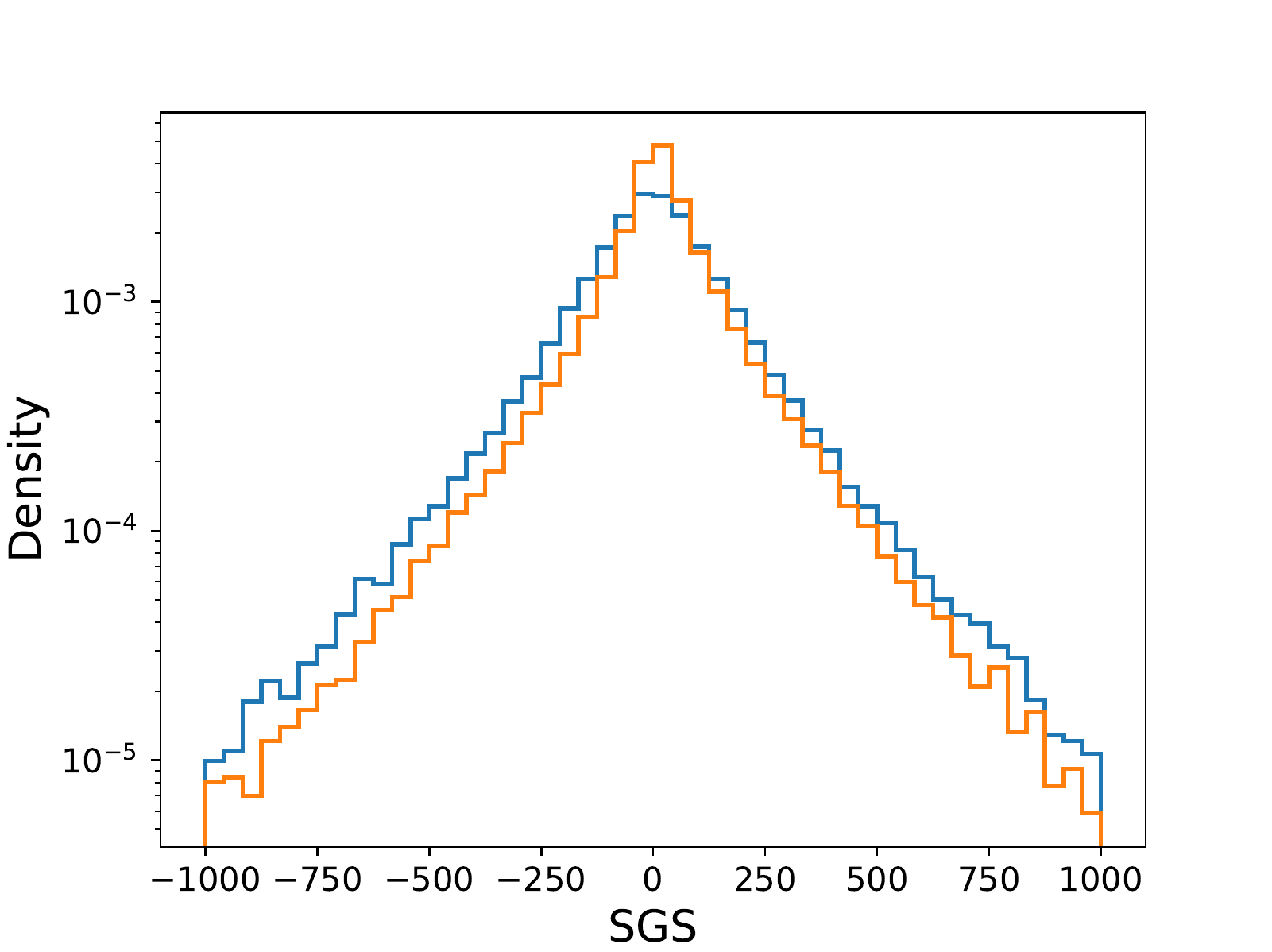}}
}
\caption{A-priori results for the probability density distributions of the true and framework predicted LES source terms for $Re=32000$ (left) and $Re=64000$ (right). Note that the training data was generated for $Re=32000$ only and prediction on $Re=64000$ represents a stringent validation. }
\label{fig:fig5}
\end{figure}

In contrast, Figure \ref{fig:fig5} shows the performance of the proposed framework in predicting sub-grid contributions purely through the indirect exposure to supervised data in the training process. The figure shows a remarkable ability for $\Pi$ reconstruction for both $Re$ values of 32000 and 64000, solely from grid-resolved quantities. Performance similar to ideal model-coefficients mentioned in the previous figure are also observed. The $Re=64000$ case is utilized to assess model performance for `out-of-training' snapshot data in an a-priori sense. The trained framework is seen to lead to viable results for a completely unseen data set with more energetic physics. We may thus conclude that the map has managed to embed a relationship between sharp spectral cutoff filtered quantities and sub-grid source terms.

\begin{figure}
\centering
\mbox{
\subfigure[$Re=32000$]{\includegraphics[width=0.9\textwidth]{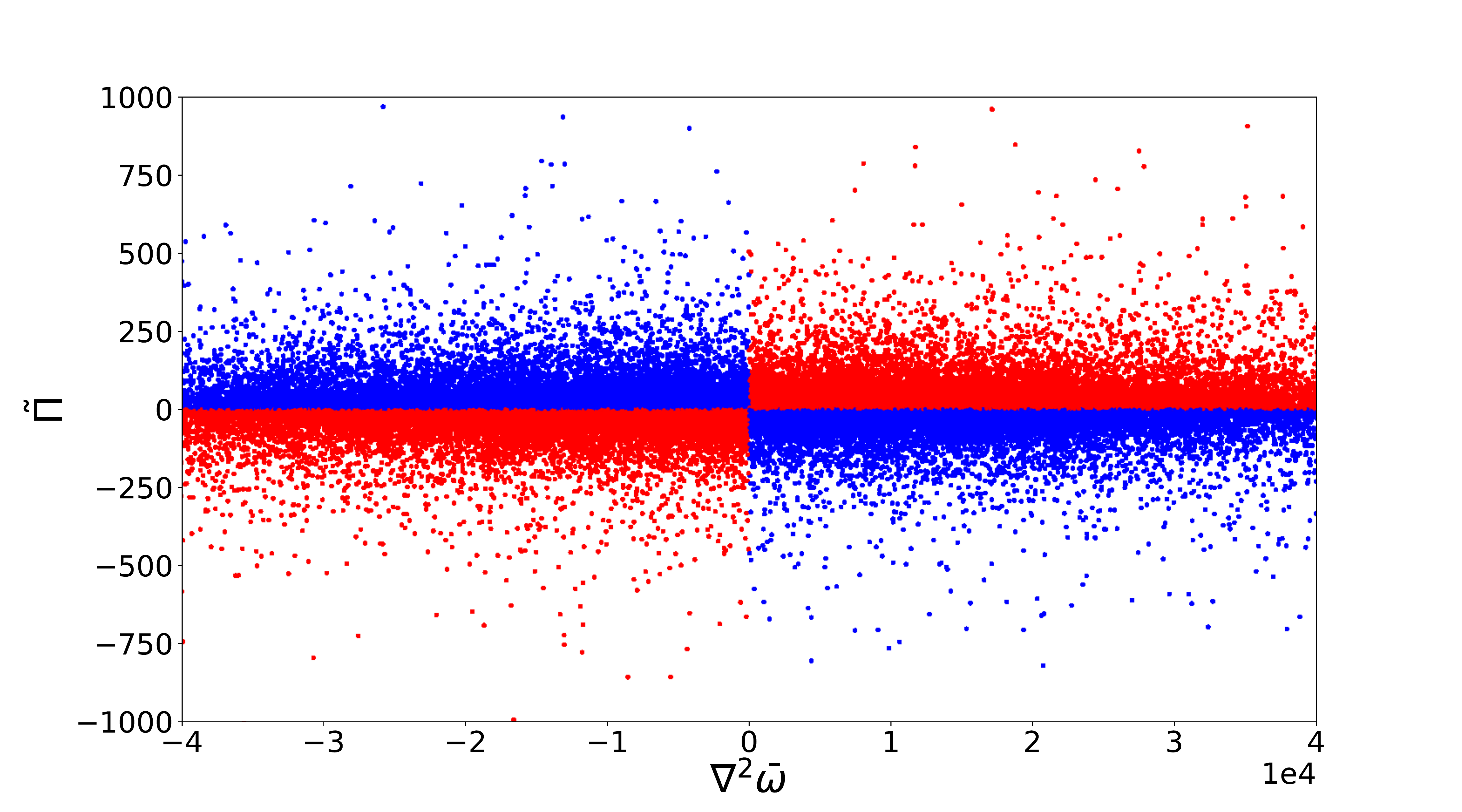}}}
\\
\mbox{
\subfigure[$Re=64000$]{\includegraphics[width=0.9\textwidth]{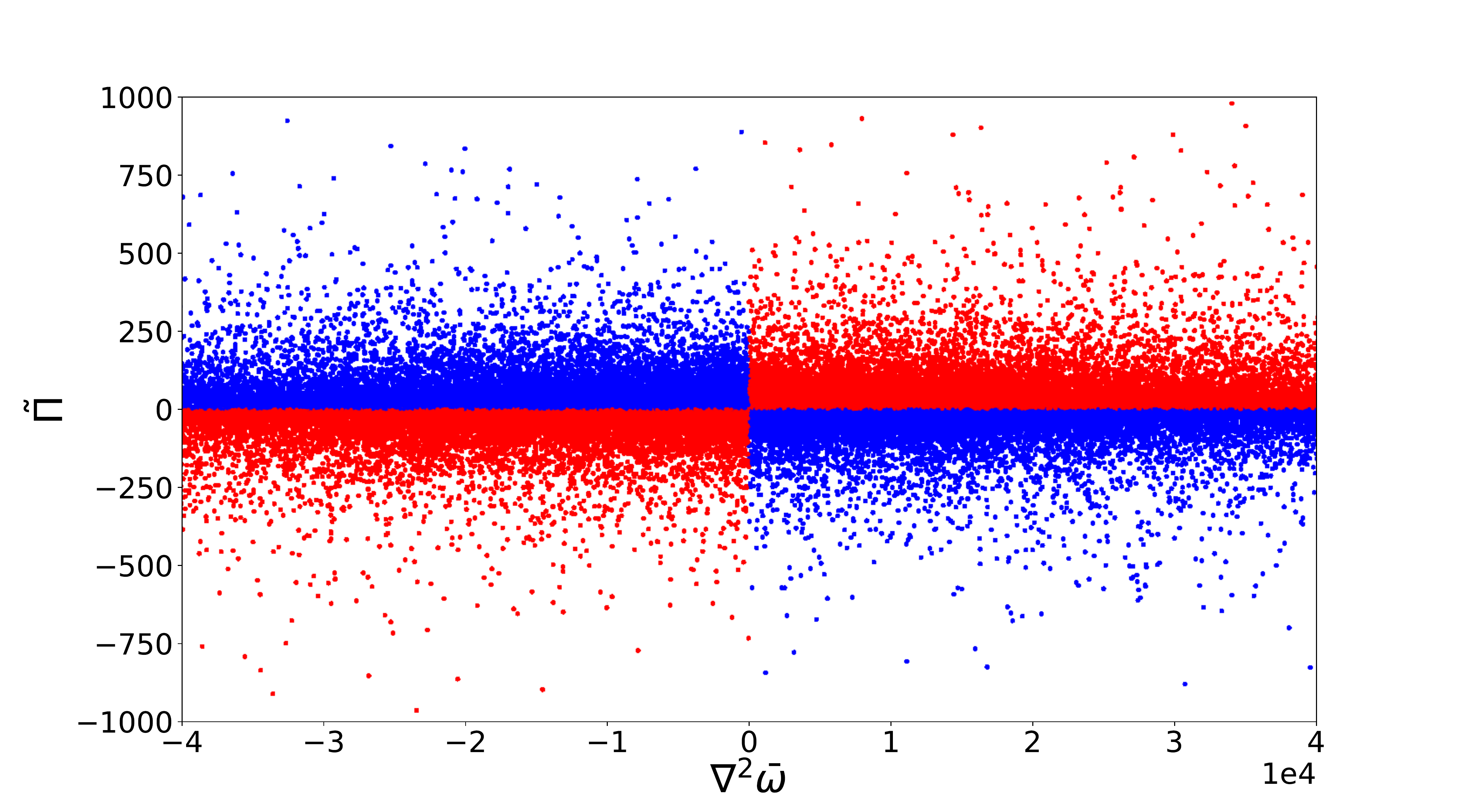}}
}
\caption{An a-priori assessment of the nature of truncation given by Equation \ref{eq9} for $t=2$ snapshot data at $Re=32000$ (top) and $Re=64000$ (bottom). The nature of this truncation is for the preservation of viscous stability in a coarse-grained forward simulation.}
\label{fig:fig6}
\end{figure}

We also visually quantify the effect of Equation \ref{eq9} (described for the process of numerical realizability) in Figure \ref{fig:fig6} where a hardwired truncation is utilized for precluding violation of viscous stability in the forward simulations of our learning deployment. One can observe that the blue regions of the figure, which are spatial locations of sub-grid forcing ($\tilde{\Pi}$) and Laplacian $\nabla^2 \bar{\omega}$ being the opposite sign, are truncated. However, we must clarify that this does not imply a constraint on the nature of forcing being obtained by our model - a negative value of the sub-grid term implies a damping of vorticity and the finer scales whereas a positive value implies production at the finer scales. Our next step is to assess the ability of this relationship to recover statistical trends in an a-posteriori deployment. The fact that roughly half of the predicted sub-grid terms are truncated matches the observations in \cite{piomelli1991subgrid} where it is observed that forward and backscatter are present in approximately equal amounts when extracted from DNS data. Studies are underway to extend some form of dynamic localization of backscatter to the current formulation along the lines of \cite{ghosal1995dynamic}.

\section{Deployment and a-posteriori assessment}

The ultimate test of any data-driven closure model is in an a-posteriori framework with subsequent assessment for the said model's ability to preserve coherent structures and scaling laws. While the authors have undertaken a-priori studies with promising results for data-driven ideologies for LES \citep{maulik2017neural}, the results of the following section are unique in that they represent a model-free turbulence model computation in temporally and spatially dynamic fashion. This test setup is particulary challenging due to the neglected effects of numerics in the a-priori training and assessment. In the following we utilize angle-averaged kinetic energy spectra to assess the ability of the proposed framework to preserve integral and inertial range statistics. In brief, we mention that the numerical implementation of the conservation laws are through second-order discretizations for all spatial quantities (with a kinetic-energy conserving Arakawa discretization \citep{arakawa1966computational} for the calculation of the nonlinear Jacobian). A third-order total-variation-diminishing Runge-Kutta method is utilized for the vorticity evolution and a spectrally-accurate Poisson solver is utilized for updating streamfunction values from the vorticity. Our proposed framework is deployed pointwise for approximate $\Pi$ at each explicit time-step until the final time of $t=4$ is reached. The robustness of the network to the effects of numerics is thus examined.

\begin{figure}
\centering
\mbox{
\subfigure[$Re=32000$]{\includegraphics[width=0.44\textwidth]{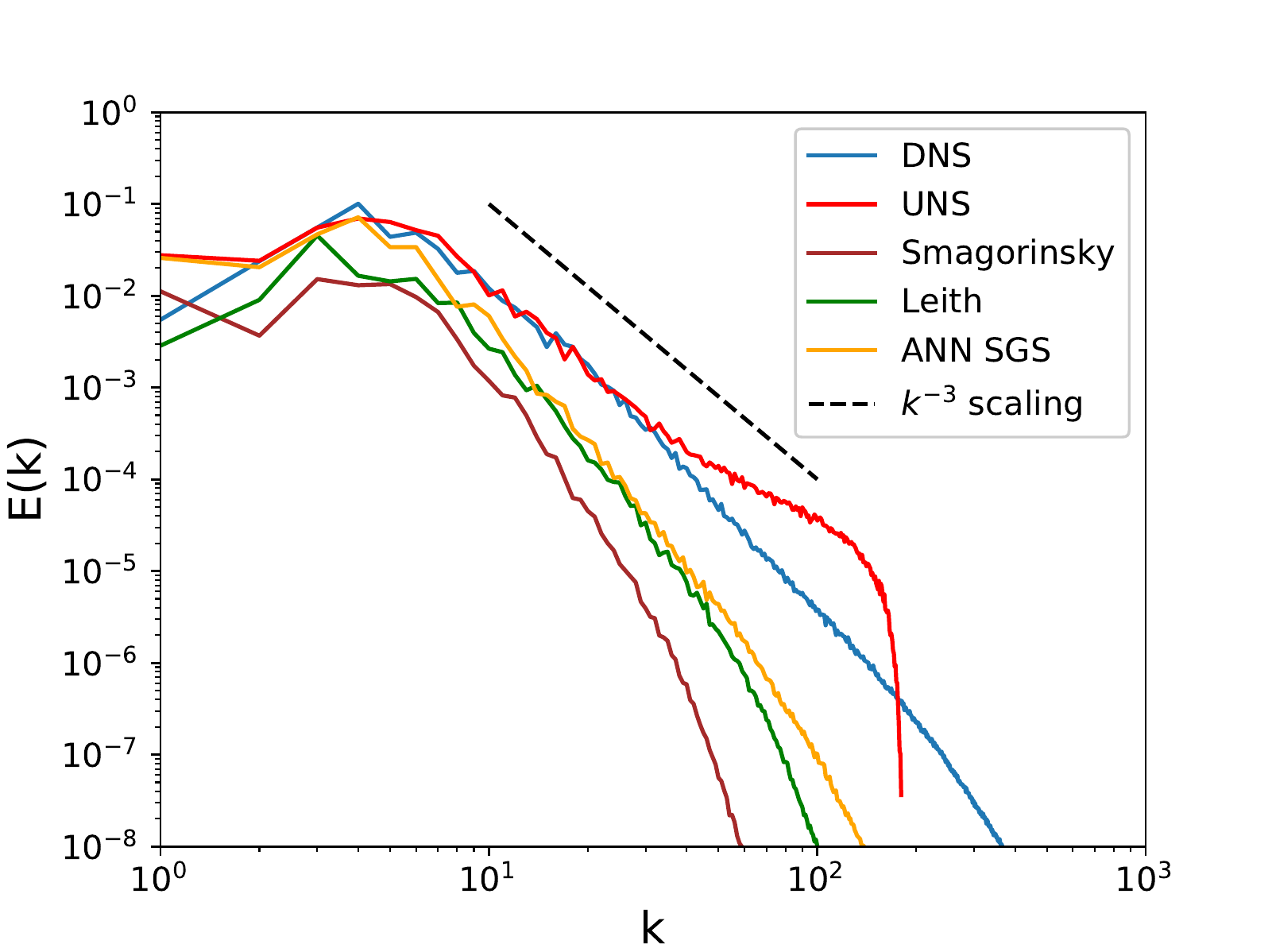}}
\subfigure[$Re=64000$]{\includegraphics[width=0.44\textwidth]{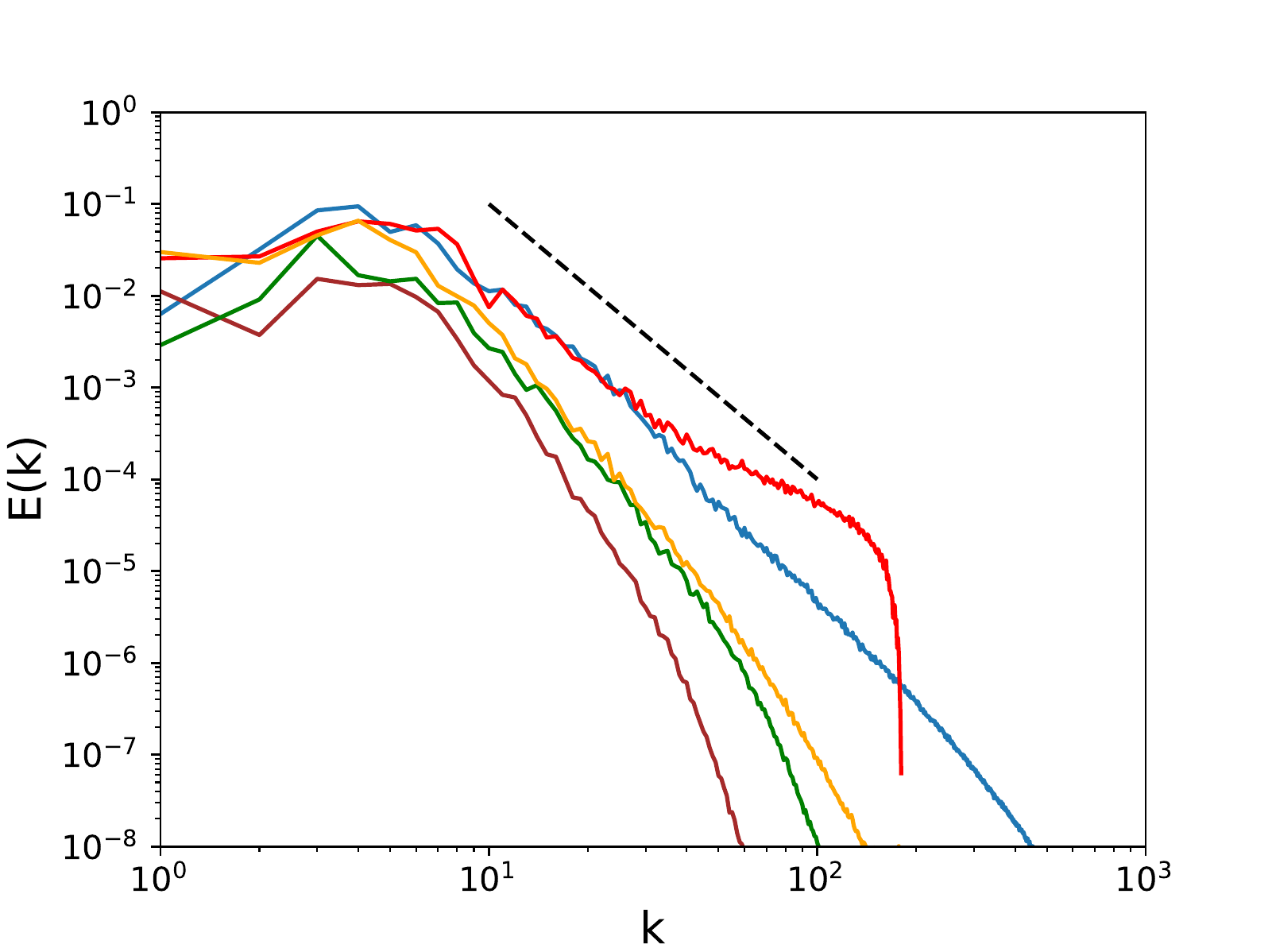}}
}
\caption{A-posteriori results for the spatially-averaged kinetic energy spectra for the proposed framework compared with DNS and UNS solutions. Note that only $Re=32000$ training data is used for both deployments and network is applied spatially and temporally in a dynamic manner until $t=4$.}
\label{fig:fig7}
\end{figure}

\begin{figure}
\centering
\mbox{
\subfigure[$Re=32000$]{\includegraphics[width=0.44\textwidth]{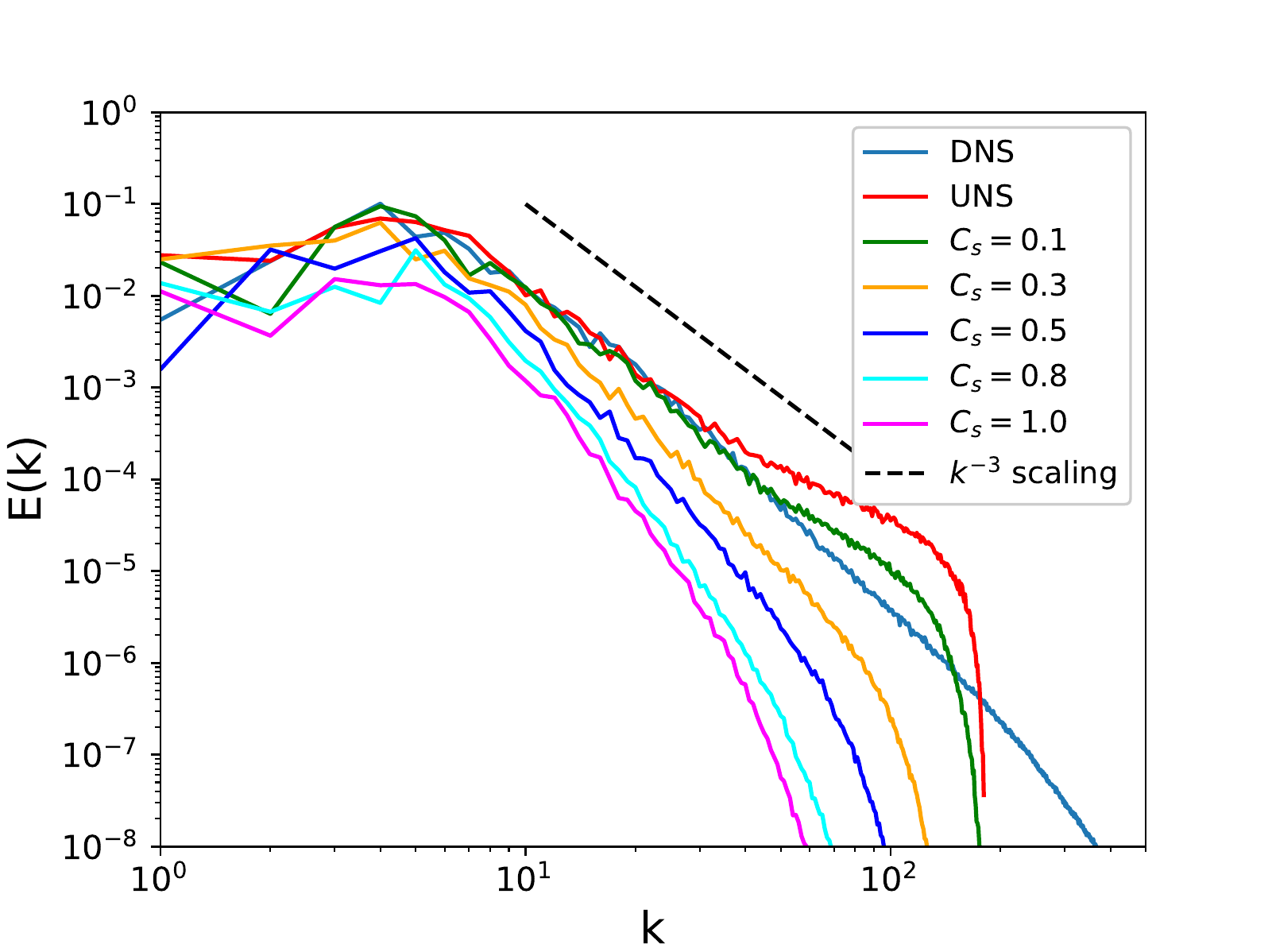}}
\subfigure[$Re=64000$]{\includegraphics[width=0.44\textwidth]{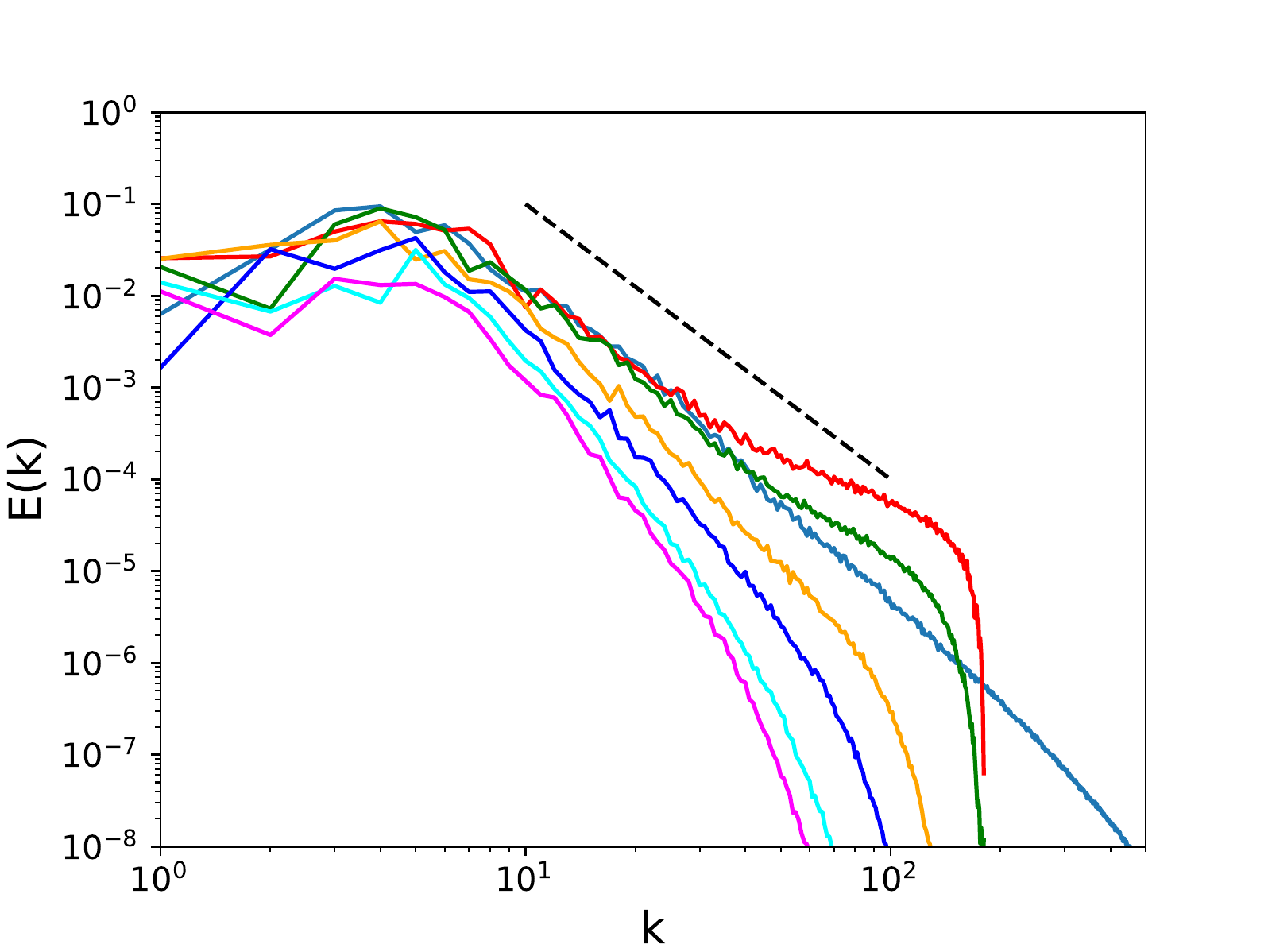}}
}
\caption{A-posteriori results for the spatially-averaged kinetic energy spectra for the Smagorinsky model for different values of their eddy-viscosity coefficients and for different Reynolds numbers at $t=4$. One can observe that the capture of lower-wavenumber energy and scaling is heavily dependant on the value of these coefficients.}
\label{fig:fig7b}
\end{figure}

\begin{figure}
\centering
\mbox{
\subfigure[$Re=32000$]{\includegraphics[width=0.44\textwidth]{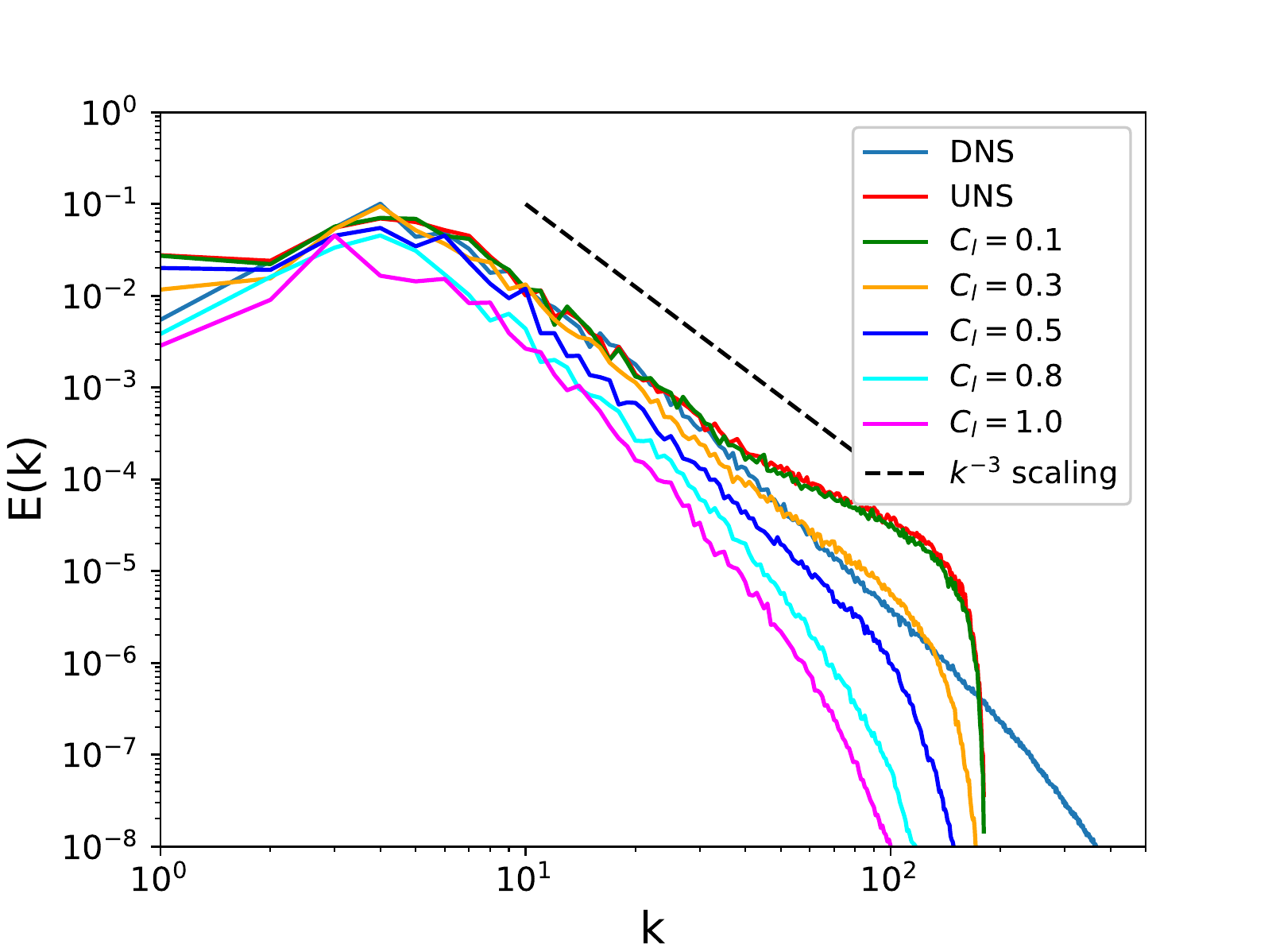}}
\subfigure[$Re=64000$]{\includegraphics[width=0.44\textwidth]{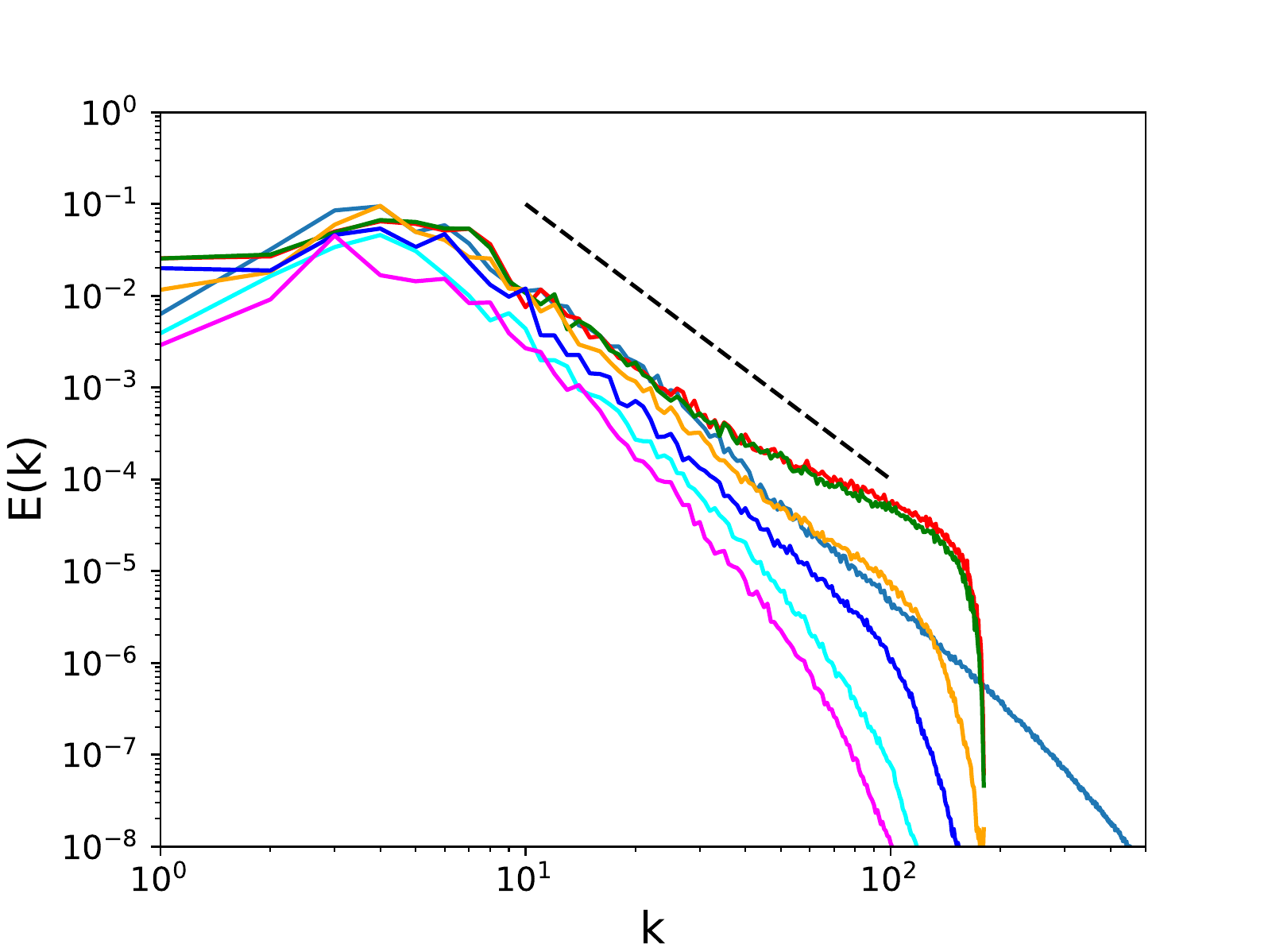}}
}
\caption{A-posteriori results for the spatially-averaged kinetic energy spectra for the Leith model for different values of their eddy-viscosity coefficients and for different Reynolds numbers at $t=4$. One can observe that the capture of lower-wavenumber energy and scaling is heavily dependant on the value of these coefficients.}
\label{fig:fig7c}
\end{figure}

Figure \ref{fig:fig7} displays the statistical fidelity of coarse-grained simulations obtained with the deployment of the proposed framework for $Re=32000$. Stable realizations of the vorticity field are generated due to the combination of our training and post-processing. For the purpose of comparison, we also include coarse-grained no-model simulations, i.e., unresolved numerical simulations (UNS) which demonstrate an expected accumulation of noise at grid cut-off wavenumbers. DNS spectra are also provided showing agreement with the $k^{-3}$ theoretical scaling expected for two-dimensional turbulence. Our proposed framework is effective at stabilizing the coarse-grained flow by estimating the effect of sub-grid quantities and preserving trends with regards to the inertial range scaling. We also demonstrate the utility of our learned map on an a-posteriori simulation for $Re=64000$ data where similar trends are recovered. This also demonstrates an additional stringent validation of the data-driven model for ensuring generalized-learning. The reader may observe that Smagorinsky and Leith turbulence model predictions using static model coefficients of value 1.0 (i.e., $C_s=C_l=1.0$) lead to over-dissipative results particularly at the lower (integral) wavenumbers. This trend is unsurprising, since the test case examined here represents non-stationary decaying turbulence for which fixed values of the coefficients are not recommended. Indeed, the application of the Smagorinsky model to various engineering and geophysical flow problems has revealed that the constant is not single-valued and varies depending on resolution and flow characteristics \citep{galperin1993large,canuto1997determination,vorobev2008smagorinsky} with higher values specifically for geophysical flows \citep{cushman2011introduction}. In comparison, the proposed framework has embedded the adaptive nature of dissipation into its map which is a promising outcome. Figures \ref{fig:fig7b} and \ref{fig:fig7c} show the performance of the Smagorinsky and Leith models, respectively, for a $Re=32000$ and $Re=64000$ a-posteriori deployment for different values of the eddy-viscosity coefficients. One can observe that the choice of the model-form coefficient is critical in the capture of the lower wavenumber fidelity.

\begin{figure}
\centering
\mbox{
\subfigure[$Re=32000$]{\includegraphics[width=0.44\textwidth]{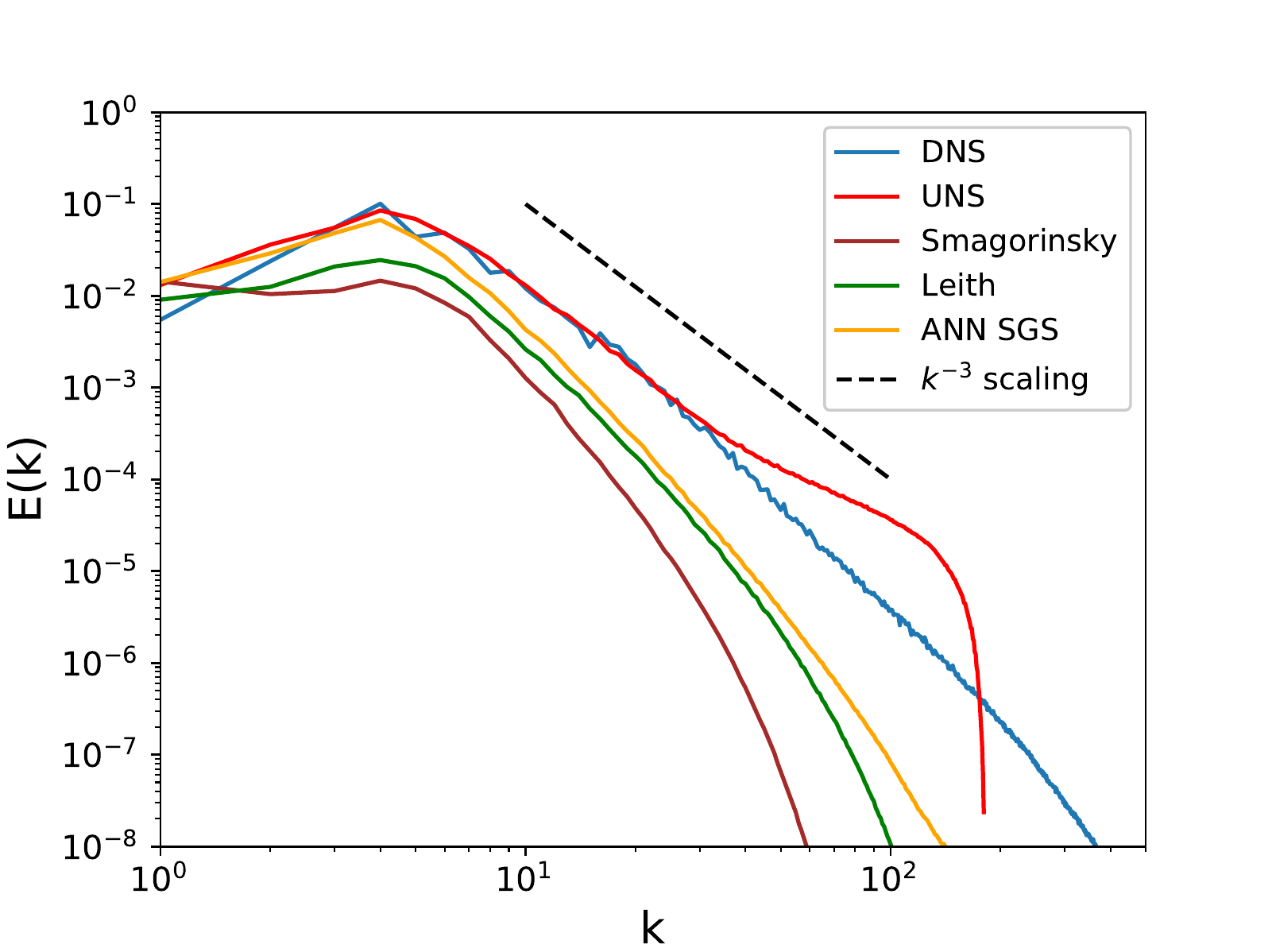}}
\subfigure[$Re=64000$]{\includegraphics[width=0.44\textwidth]{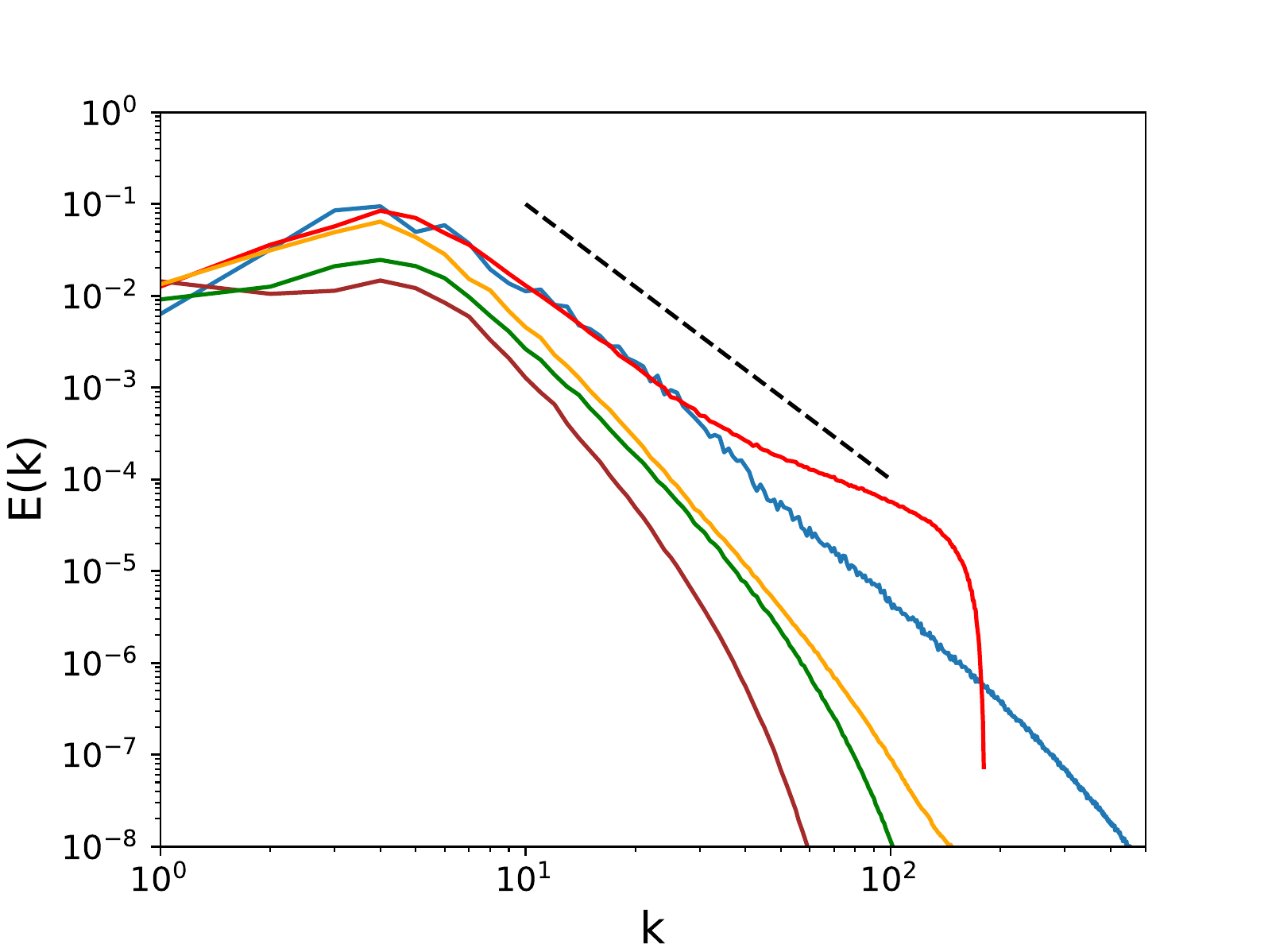}}
}
\caption{A-posteriori results for 24 ensemble-averaged simulations for $Re=32000$ (left) and $Re=64000$ (right).}
\label{fig:fig8}
\end{figure}

\begin{figure}
\centering
\mbox{
\subfigure[$Re=32000$]{\includegraphics[width=0.44\textwidth]{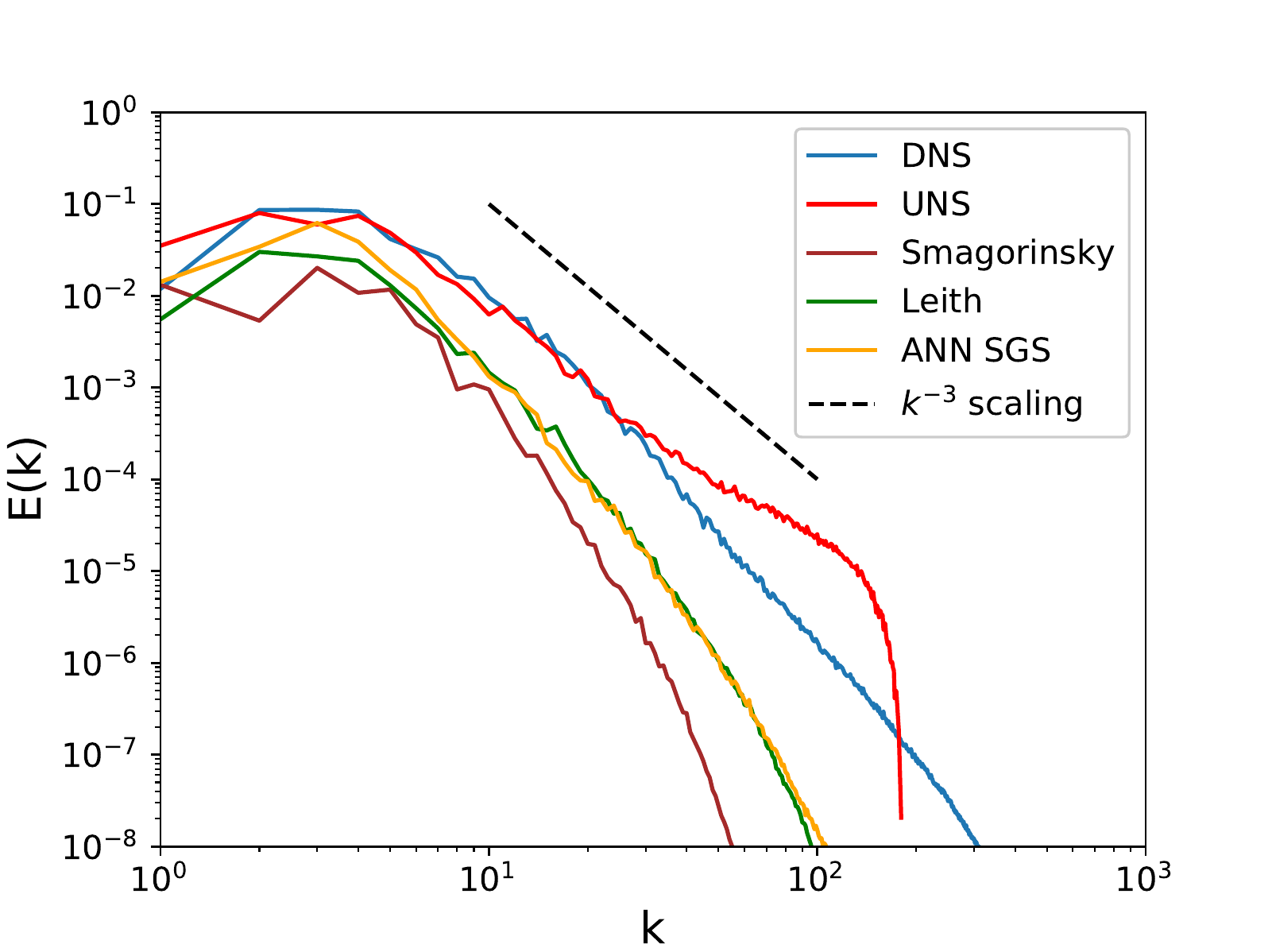}}
\subfigure[$Re=64000$]{\includegraphics[width=0.44\textwidth]{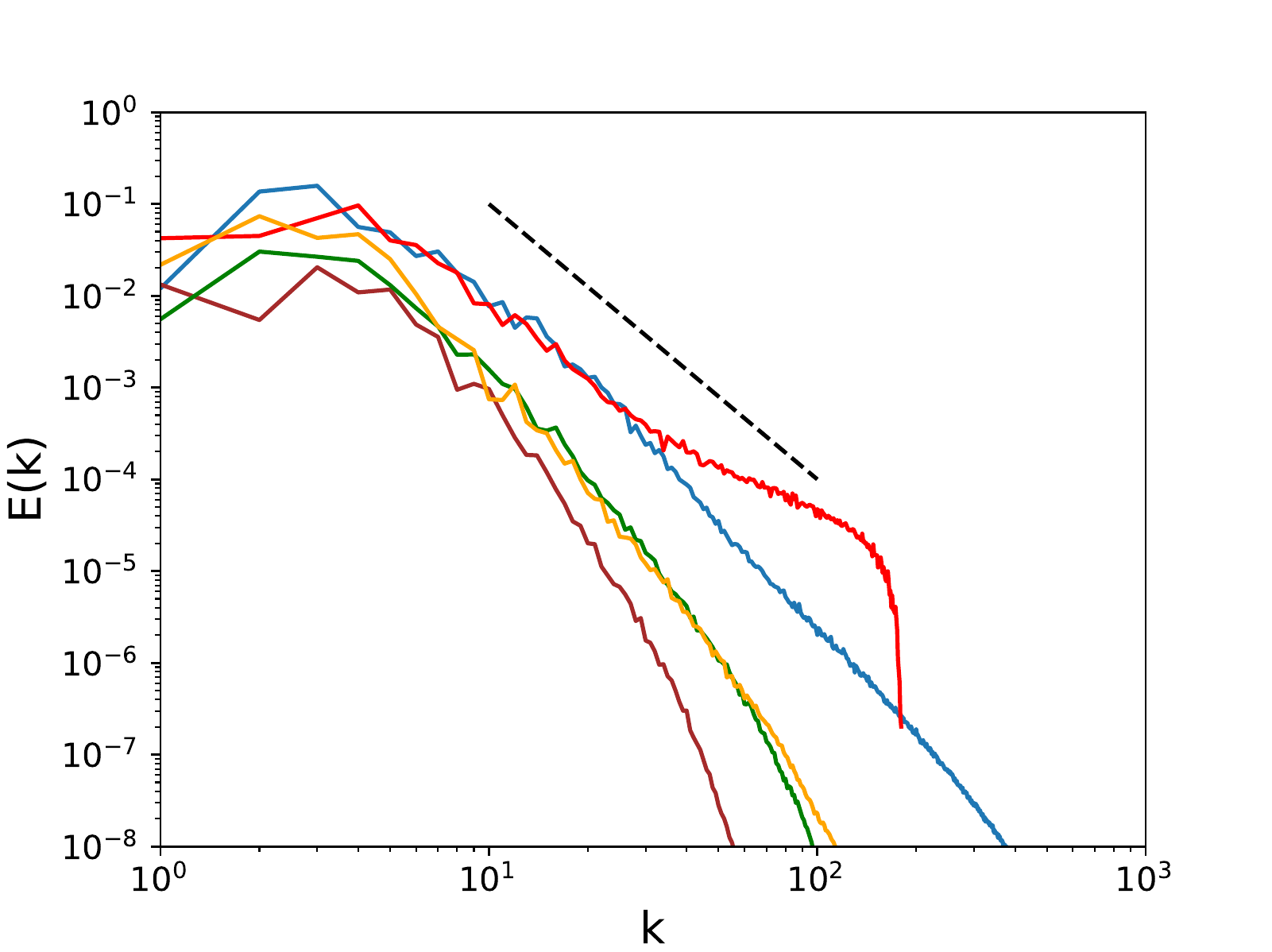}}
}
\caption{The deployment of our framework till $t=6$ for $Re=32000$ (left) and $Re=64000$ (right) showing that a sub-grid model has been learned for utility beyond the training region. We note that the training region is defined between $t=0$ and $t=4$ alone.}
\label{fig:fig9}
\end{figure}

In particular, we would like to note that the choice of a coarse-grained forward simulation using a Reynolds number of $64000$ represents a test for establishing what the model has learned. This forward simulation verifies if the closure performance of the framework is generalizable and not a numerical artifact. A similar performance of the model on a different deployment scenario establishes the hybrid nature of our framework where the bulk behavior of the governing law is retained (through the vorticity-streamfunction formulation) and the artificial intelligence acts as a corrector for statistical fidelity. This observation holds promise for the development of closures which are generalizable to multiple classes of flow without being restricted by initial or boundary conditions. To test the premise of this hypothesis, we also display ensemble-averaged kinetic energy spectra from multiple coarse-grained simulations at $Re=32000$ and at $Re=64000$, utilizing a different set of random initial conditions for each test case. In particular, we utilize 24 different tests for averaged spectra which are displayed in Figure \ref{fig:fig8}. We would like to emphasize here that the different initial conditions correspond to the same initial energy spectrum in wavenumber space but with random vorticity fields in Cartesian space. The performance of our proposed framework is seen to be repeatable across different instances of random initial vorticity fields sharing the same energy spectra. Details related to the generation of these random initial conditions may be found in \cite{maulik2017stable}. In addition, we also display spectra obtained from an a-posteriori deployment of our framework till $t=6$ for $Re=32000$ and $Re=64000$ , shown in Figure \ref{fig:fig9}, which ensures that the model has learned a sub-grid closure effectively and predicts the vorticity forcing adequately in a temporal region that it has not been exposed to during training.

\begin{figure}
\centering
\mbox{
\subfigure[ANN SGS - $Re=32000$]{\includegraphics[width=0.44\textwidth]{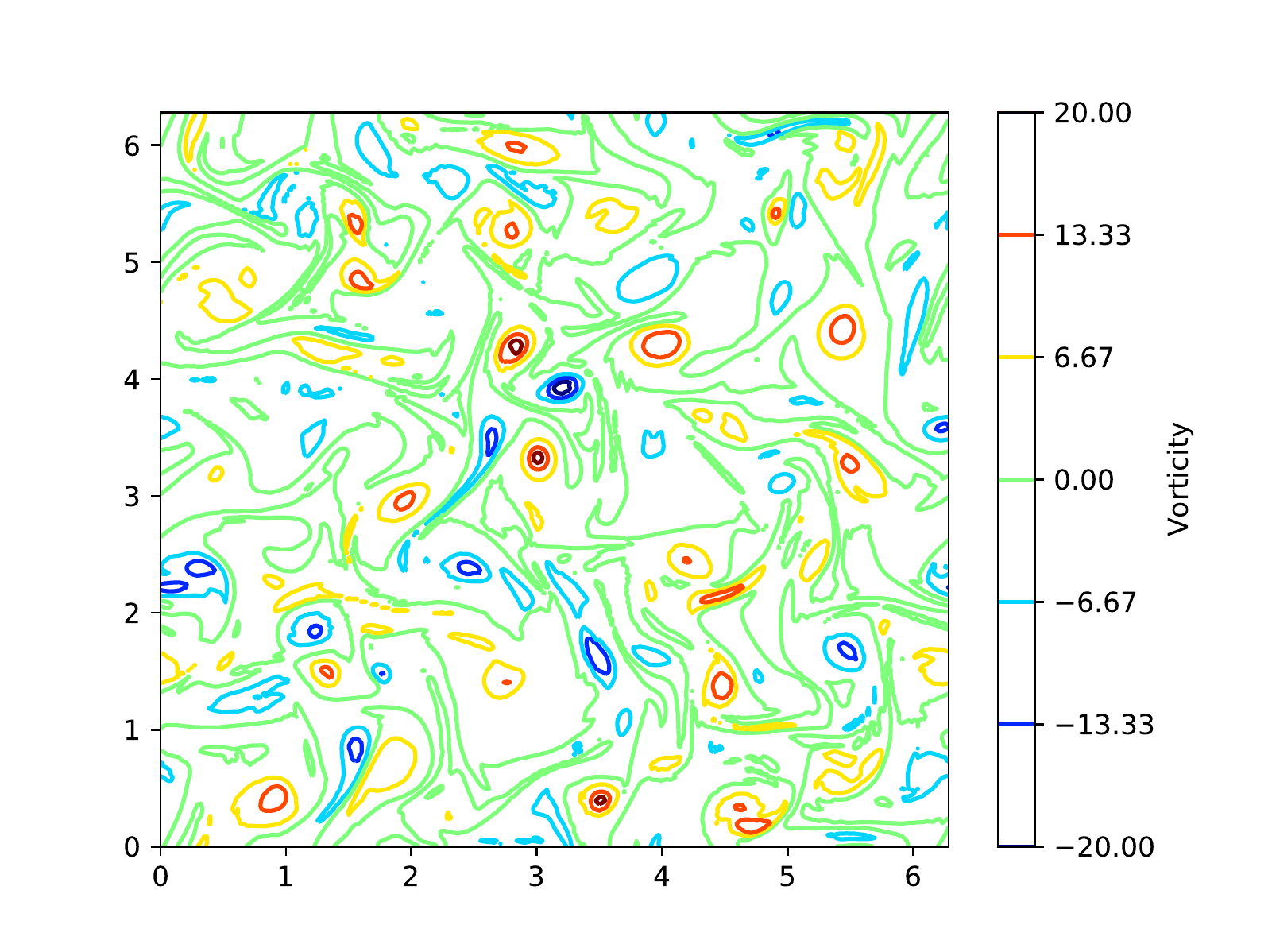}}
\subfigure[ANN SGS - $Re=64000$]{\includegraphics[width=0.44\textwidth]{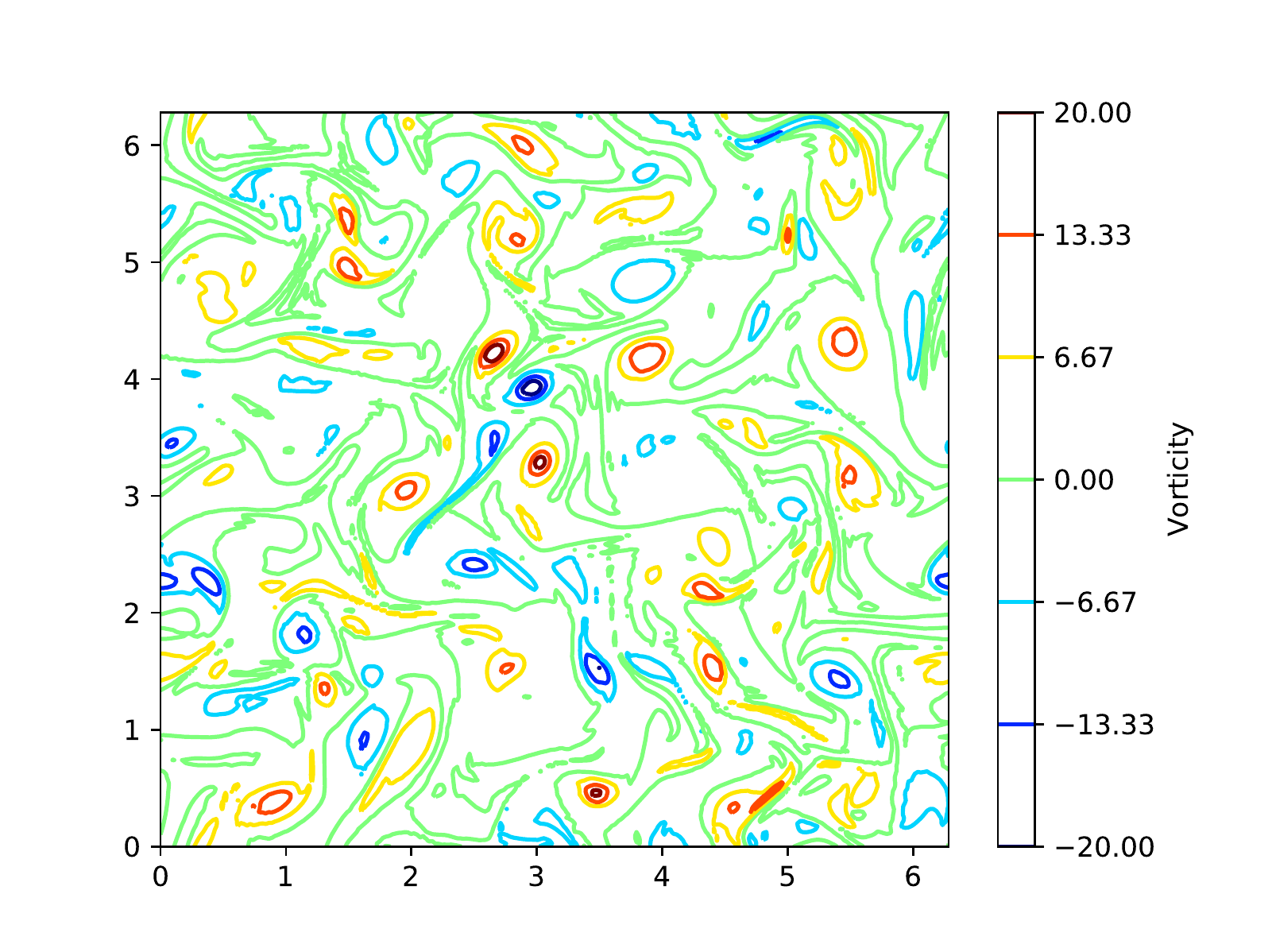}}
} \\
\mbox{
\subfigure[UNS - $Re=32000$]{\includegraphics[width=0.44\textwidth]{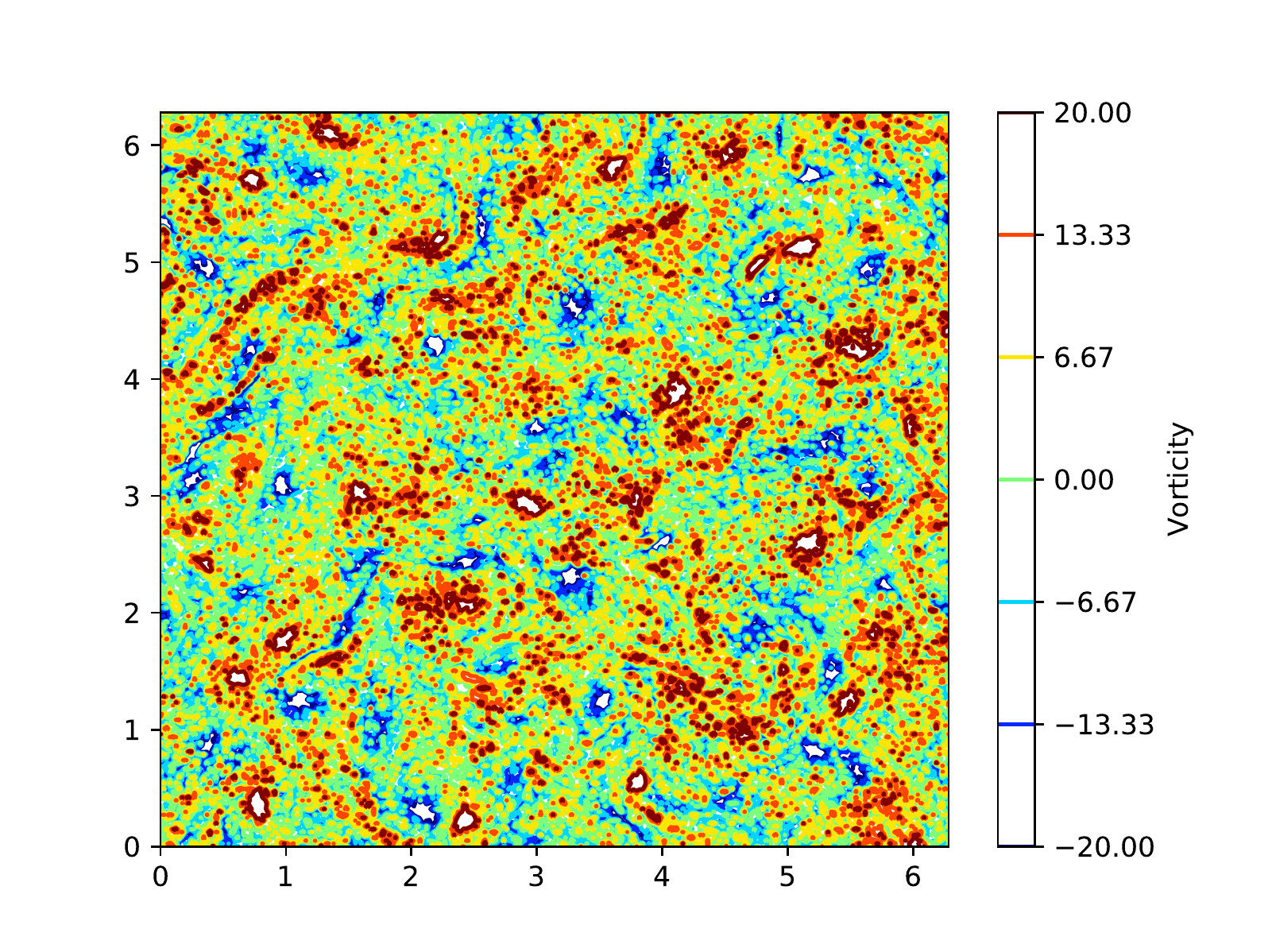}}
\subfigure[UNS - $Re=64000$]{\includegraphics[width=0.44\textwidth]{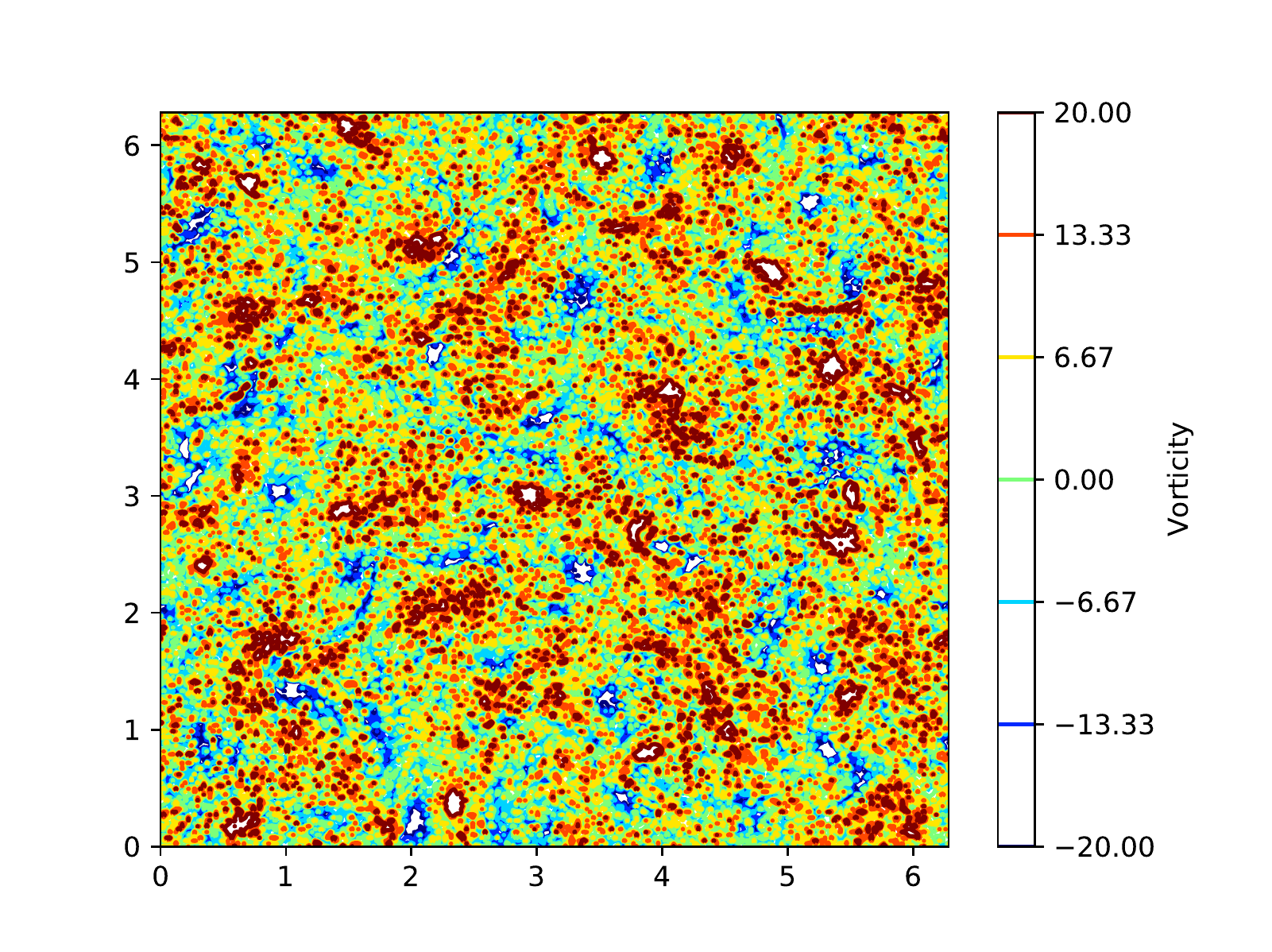}}
} \\
\mbox{
\subfigure[FDNS - $Re=32000$]{\includegraphics[width=0.44\textwidth]{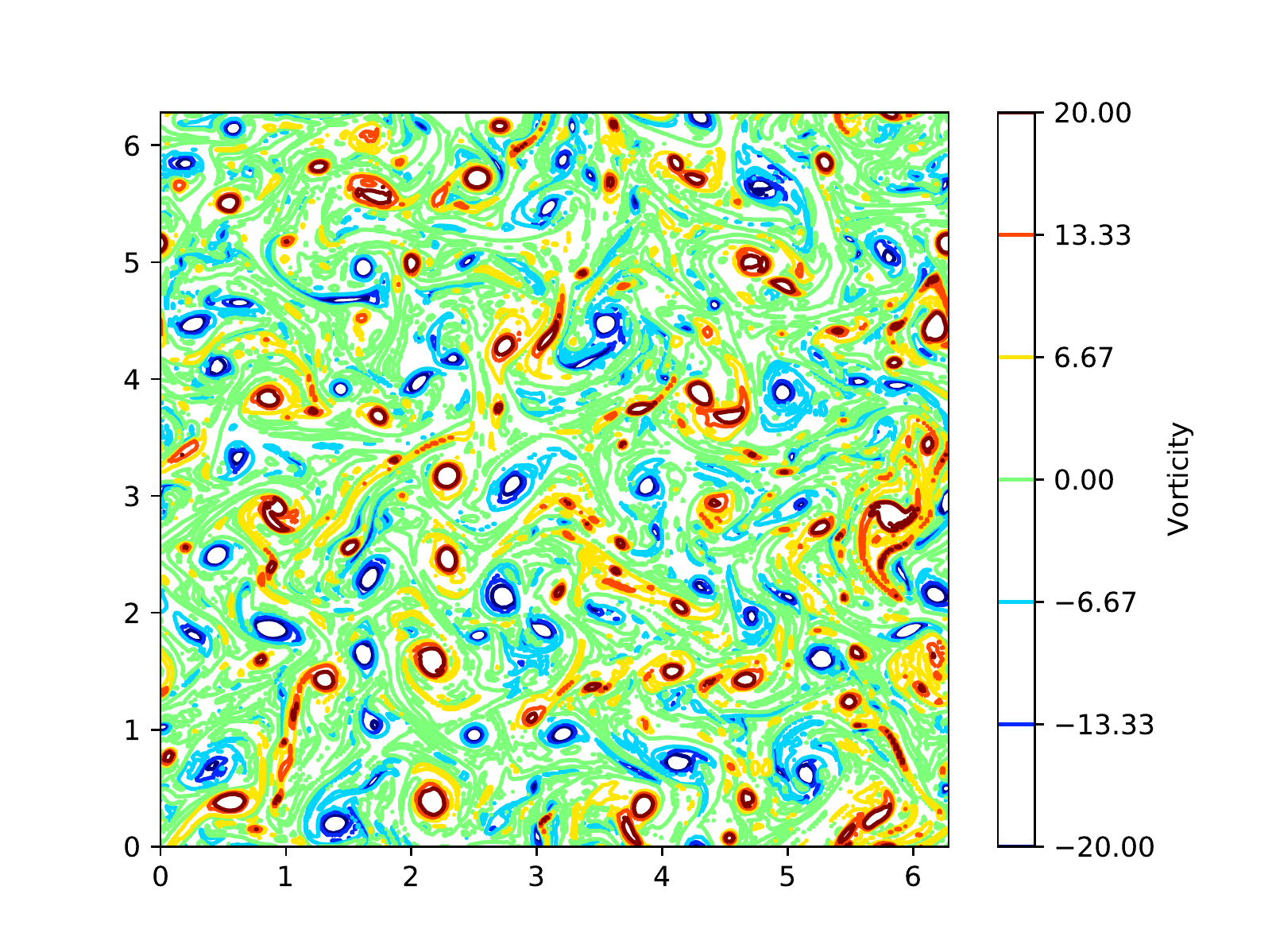}}
\subfigure[FDNS - $Re=64000$]{\includegraphics[width=0.44\textwidth]{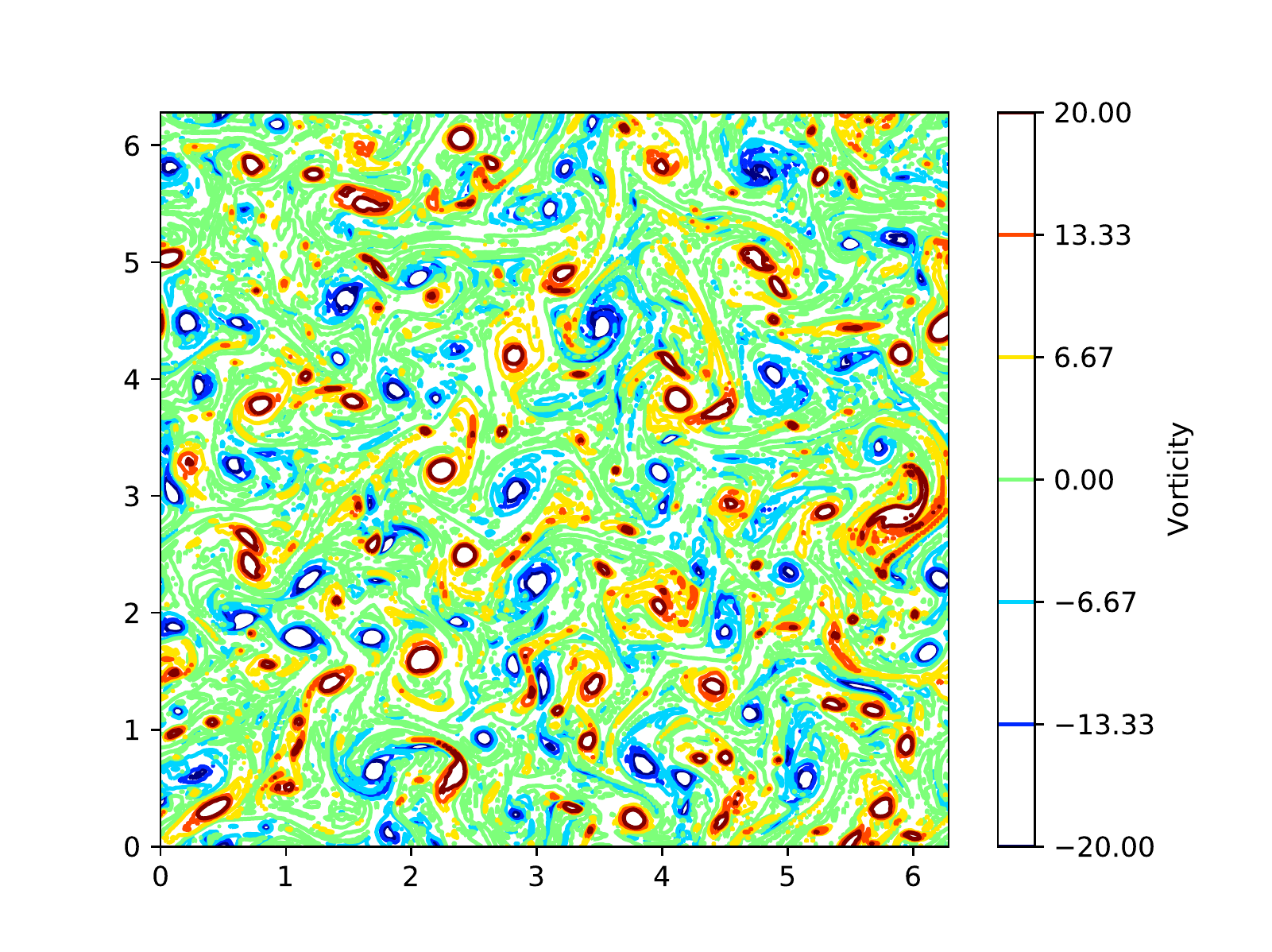}}
} \\
\caption{A-posteriori results for the proposed framework showing vorticity fields for $Re=32000$ and $Re=64000$ data using coarse-grained grids (top). We also provide no-model simulations (middle) and filtered DNS contours (bottom) for the purpose of comparison. }
\label{fig:fig10}
\end{figure}

Figure \ref{fig:fig10} shows a qualitative assessment of the stabilization property of machine learning framework where a significant reduction in noise can be visually ascertained due its deployment. Coherent structures are retained successfully as against UNS results where high-wavenumber noise is seen to corrupt field realizations heavily. Filtered DNS (FDNS) data obtained by Fourier cut-off filtering of vorticity data obtained from DNS are also shown for the purpose of comparison. As discussed previously, the stabilization behavior is observed for both $Re=32000$ and $Re=64000$ data. We may thus conclude that the learned model has established an implicit sub-grid model as a function of grid-resolved variables. We reiterate that the choice of the eddy-viscosities is motivated by ensuring a fair comparison with the static Smagorinsky and Leith sub-grid models and studies are underway to increase complexity in the mapping as well as input space.

\section{A-priori and a-posteriori dichotomy}

In the previous sections, we have outlined the performance of our proposed framework according to the optimal model architecture chosen by a grid-search for the number of hidden layers as well as the number of hidden-layer neurons. This a-priori hyper-parameter selection is primarily devised on mean-squared-error minimization and is susceptible to providing model architectures which are less resistant to over-fitting and more prone to extrapolation. Our experience shows that an a-posteriori prediction (such as for this simple problem) must be embedded into the model selection decision process to ensure an accurate learning of physics. We briefly summarize our observations of the a-priori and a-posteriori dichotomy in the following.

\subsection{Effect of eddy-viscosity inputs}

By fixing our optimal set of hyper-parameters (i.e., a two-layer 50 neuron network), we attempted to train a map using an input space without the choice of Smagorinsky and Leith viscosity kernels. Therefore our inputs would simply be the 9-point stencils for vorticity and streamfunction as shown in the mathematical expression given by
\begin{align}
\label{eq13}
\begin{gathered}
\mathbb{M} : \{ \bar{\omega}_{i,j}, \bar{\omega}_{i,j+1}, \bar{\omega}_{i,j-1}, \hdots, \bar{\omega}_{i-1,j-1}, \\ \bar{\psi}_{i,j}, \bar{\psi}_{i,j+1}, \bar{\psi}_{i,j-1}, \hdots, \bar{\psi}_{i-1,j-1} \in \mathbb{R}^{18} \rightarrow \{ \tilde{\Pi}_{i,j}\} \in \mathbb{R}^1.
\end{gathered}
\end{align}

\begin{figure}
\centering
\mbox{
\subfigure[Learning rate]{\includegraphics[width=0.44\textwidth]{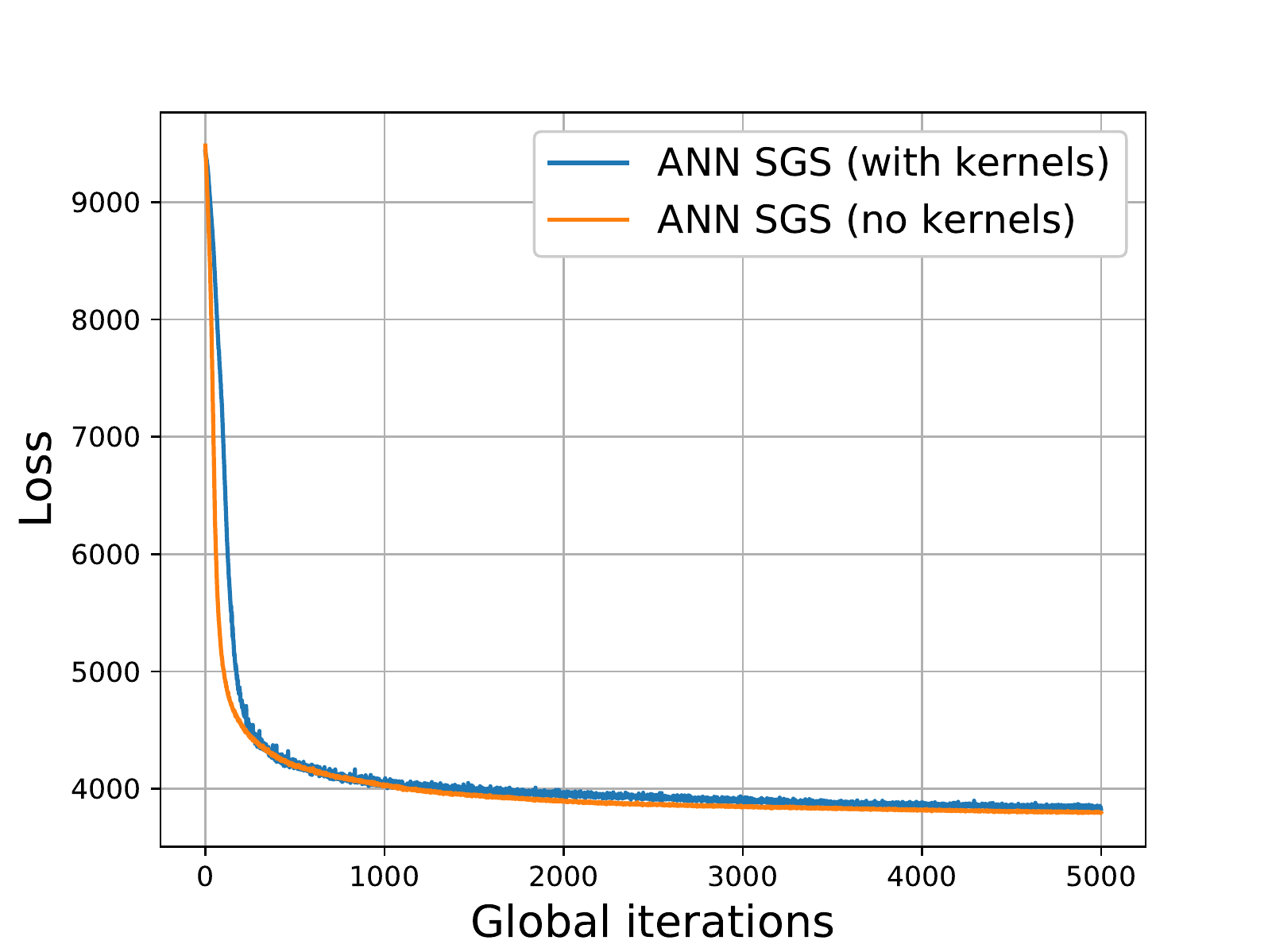}}
\subfigure[A-posteriori deployment]{\includegraphics[width=0.44\textwidth]{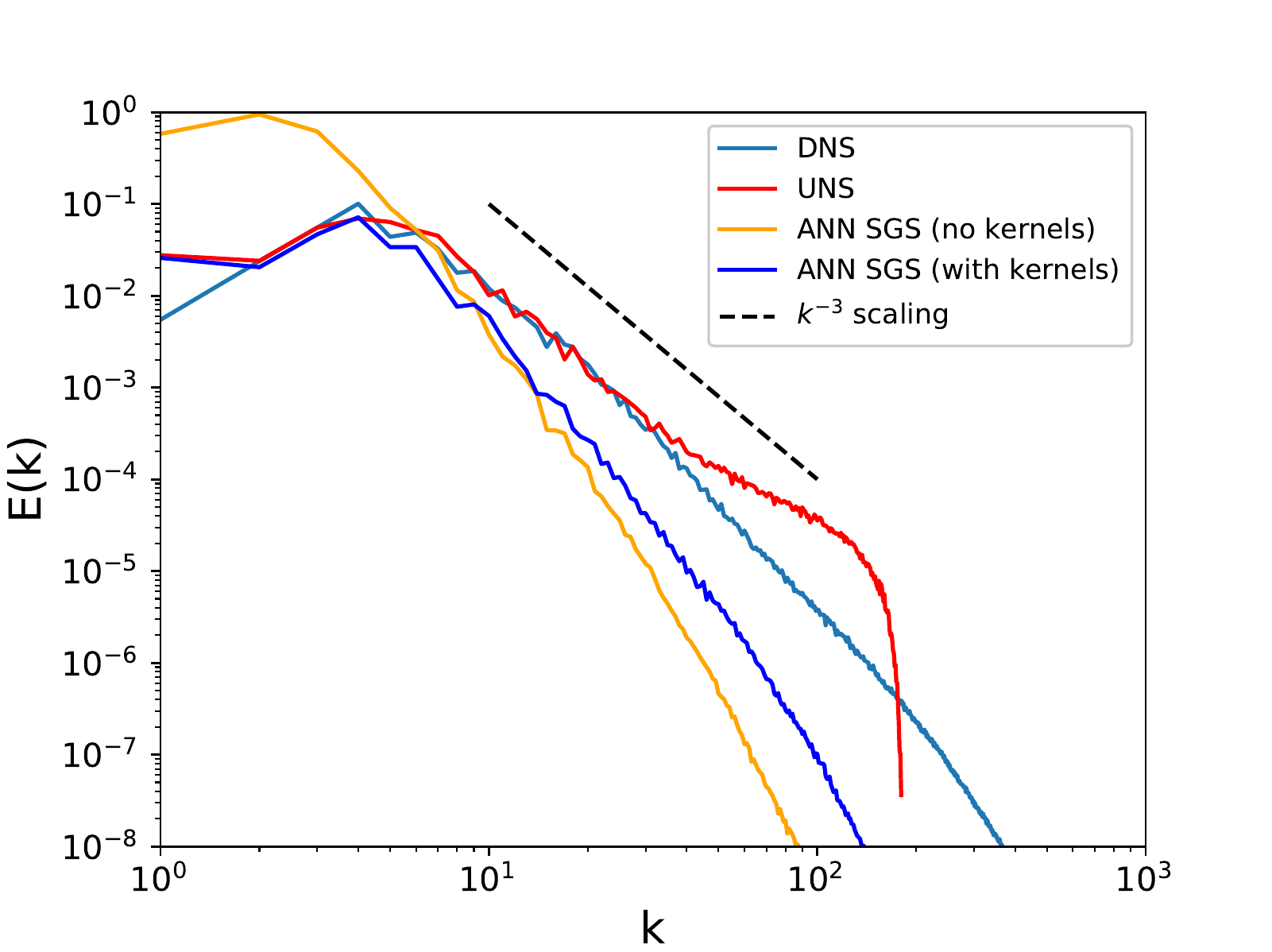}}
}
\caption{A-priori (left) and a-posteriori (right) effect of the utilization of eddy-viscosity kernel inputs in training and deployment for a two-layer 50 neuron network with a 9-point stencil. The presence of these kernels (intangible in a-priori error minimization) leads to constrained statistical fidelity in a-posteriori deployment at $Re=32000$.}
\label{fig:fig11}
\end{figure}

As shown in Figure \ref{fig:fig11}, the modification of our input space had very little effect on the training performance of our optimal network architecture. This would initially seem to suggest that the Smagorinsky and Leith kernels were not augmenting learning in any manner. However, our a-posteriori deployment of this model which mapped to sub-grid quantities from the 18-dimensional input space displayed an unconstrained behavior at the larger scales with the formation of non-physical large scale structures (also shown in Figure \ref{fig:fig8}). This strongly points towards an implicit regularization of our model due to the selection of input dimensions with these kernels.

\begin{figure}
\centering
\mbox{
\subfigure[Learning rate]{\includegraphics[width=0.44\textwidth]{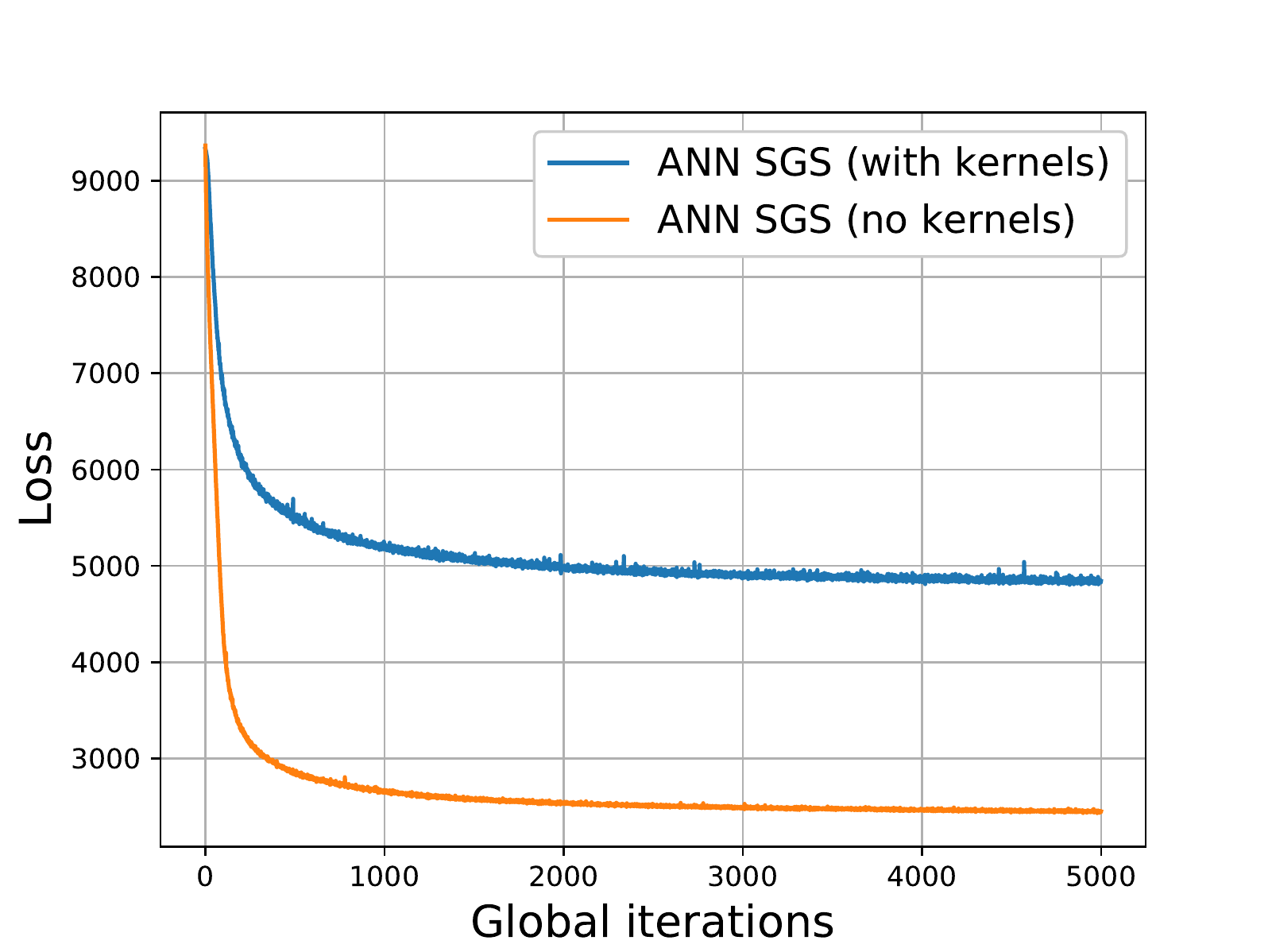}}
\subfigure[A-posteriori deployment]{\includegraphics[width=0.44\textwidth]{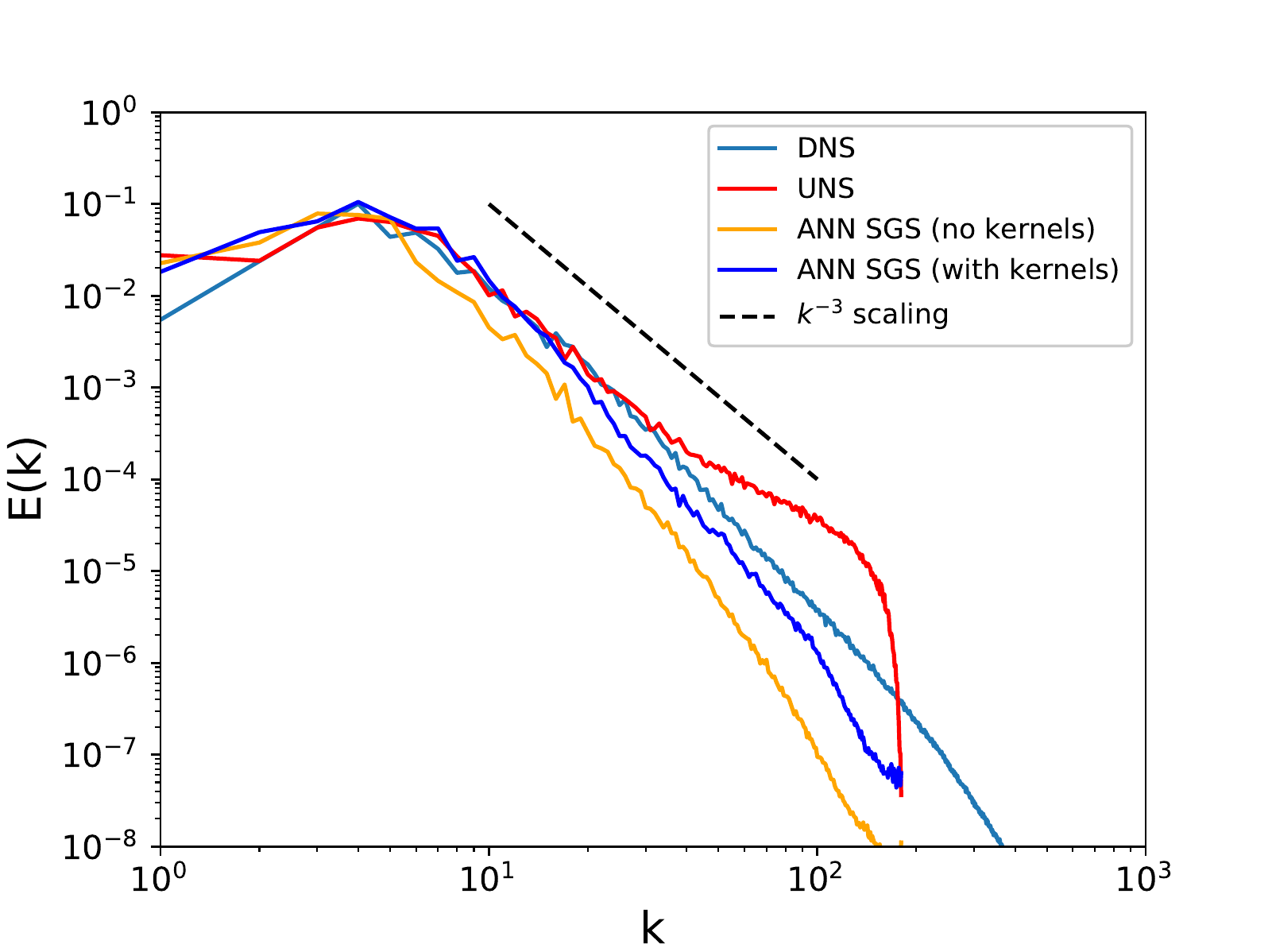}}
}
\caption{A-priori (left) and a-posteriori (right) effect of the utilization of eddy-vicsosity kernel inputs in training and deployment for a five-layer 50 neuron network with a 9-point stencil. The presence of these kernels leads to higher training errors but viable statistical fidelity in a-posteriori deployment at $Re=32000$.}
\label{fig:fig12}
\end{figure}

We undertook the same study for a 5-layer, 50 neuron ANN (one that was deemed too complex by our grid-search) with results shown in Figure \ref{fig:fig12}. Two conclusions are apparent here - the utilization of these kernels in the learning process has prevented a-priori reduction of training error at a much higher value and that the deployment of both networks (i.e., with and without input viscosities) has led to a constrained prediction of the $k^{-3}$ spectral scaling. Large scale statistical predictions remain unchanged and indeed, a better agreement with the DNS spectrum can be observed with the deeper network with the use of the kernels.

\subsection{A-posteriori informed architecture selection}

\begin{figure}
\centering
\mbox{
\subfigure[Learning rate]{\includegraphics[width=0.44\textwidth]{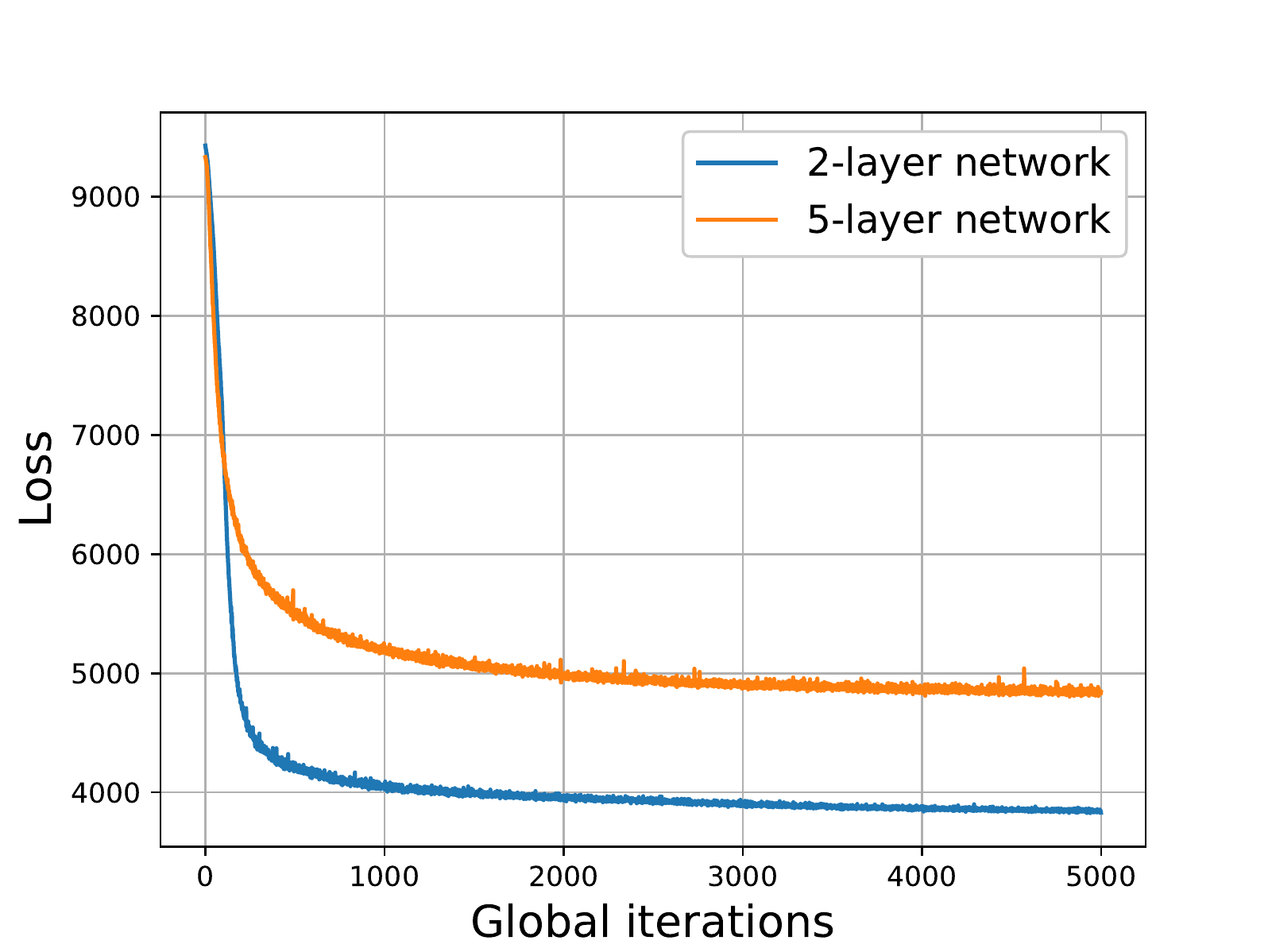}}
\subfigure[A-posteriori deployment]{\includegraphics[width=0.44\textwidth]{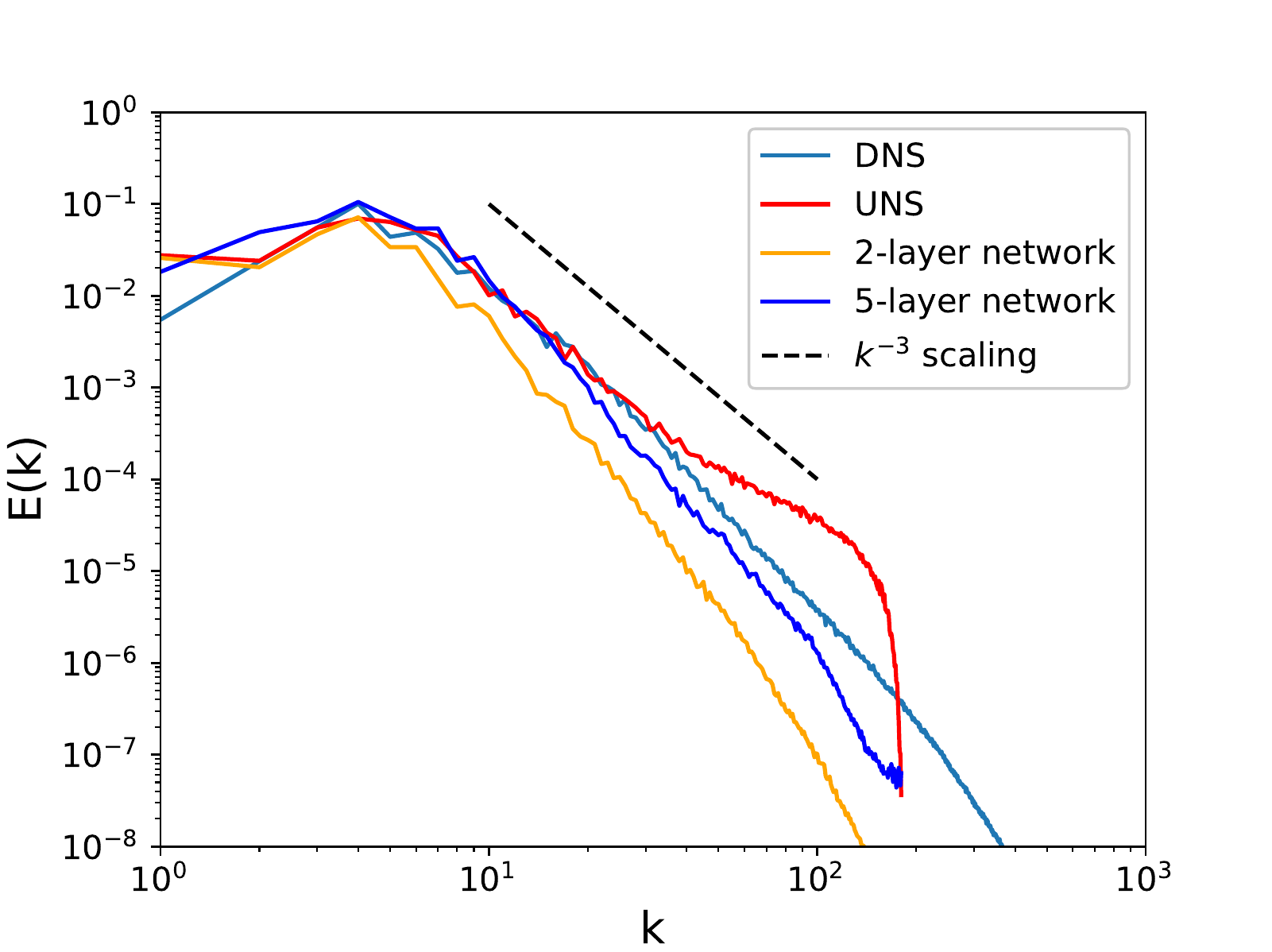}}
}
\caption{A-priori (left) and a-posteriori (right) effect of the number of hidden-layers in the proposed framework. While the two-layered ANN with a 9-point stencil leads to excellent a-priori results, the five-layered network predicts $k^{-3}$ scaling more accurately in deployment for an a-posteriori simulaion at $Re=32000$.}
\label{fig:fig13}
\end{figure}

While a-priori hyper-parameter tuning is classically utilized for most machine-learning deployments, the enforcement of physical realizability constraints (such as those given by Equation \ref{eq9}) and the presence of numerical errors during deployment may often necessitate architectures which differ significantly during a-posteriori deployment. This article demonstrates the fact that while constrained predictions are obtained by our optimal two-layer network (obtained by a grid-search), the utilization of a deeper network actually leads to more accurate predictions of the Kraichnan turbulence spectrum as shown in Figure \ref{fig:fig13}. This despite the fact that the deeper network displays a great mean-squared-error during the training phase (which was the root-cause of it being deemed ineligible in the hyper-parameter tuning). Figure \ref{fig:fig12} thus tells us that it is important to couple some form of a-posteriori analysis during model-form selection before it is deemed optimal (physically or computationally) for deployment. We note that both networks tested in this subsection utilized the Smagorinsky and Leith eddy-viscosities in their input space.

\subsection{Stencil selection}

\begin{figure}
\centering
\mbox{
\subfigure[Learning rate]{\includegraphics[width=0.44\textwidth]{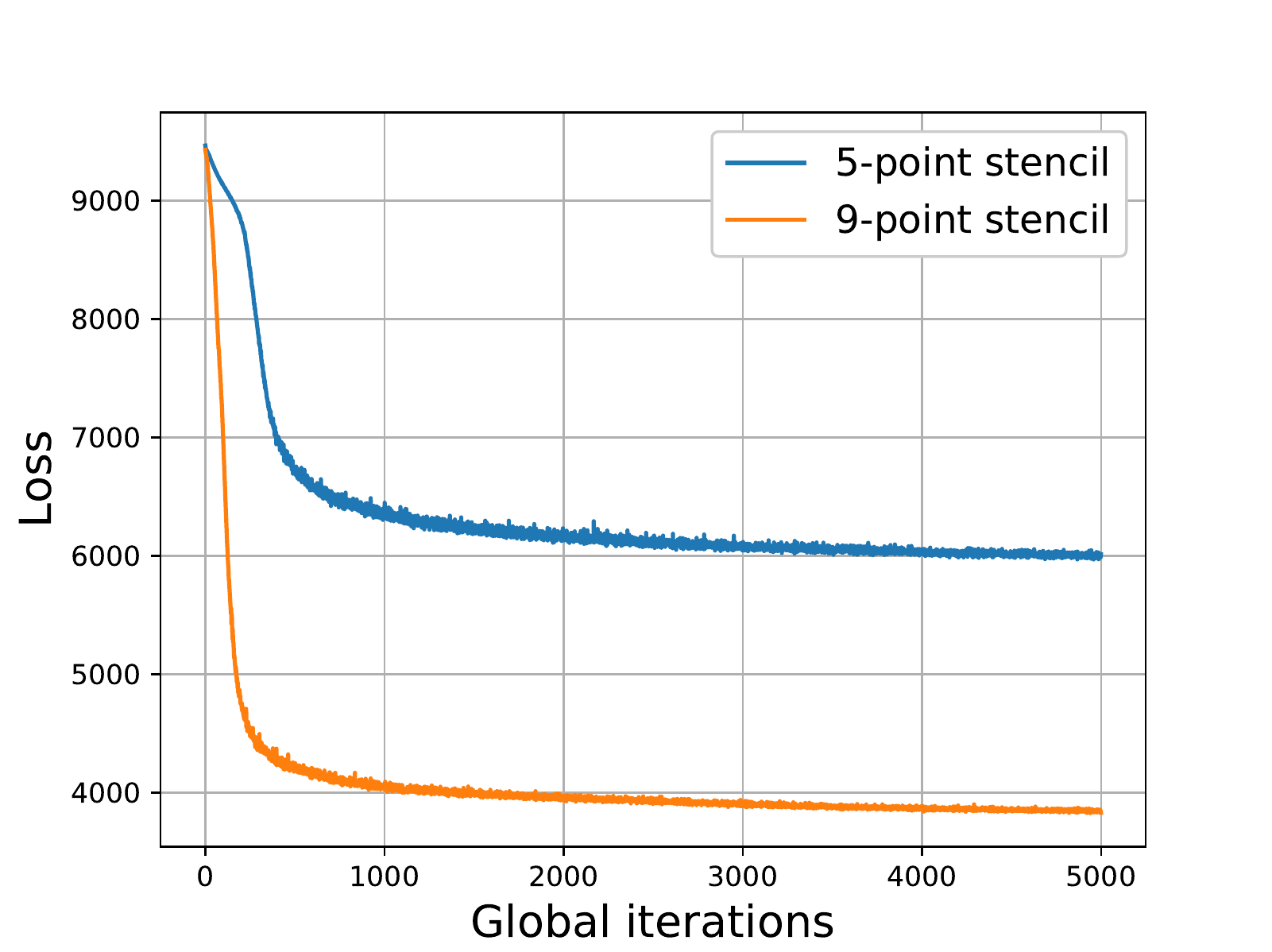}}
\subfigure[A-posteriori deployment]{\includegraphics[width=0.44\textwidth]{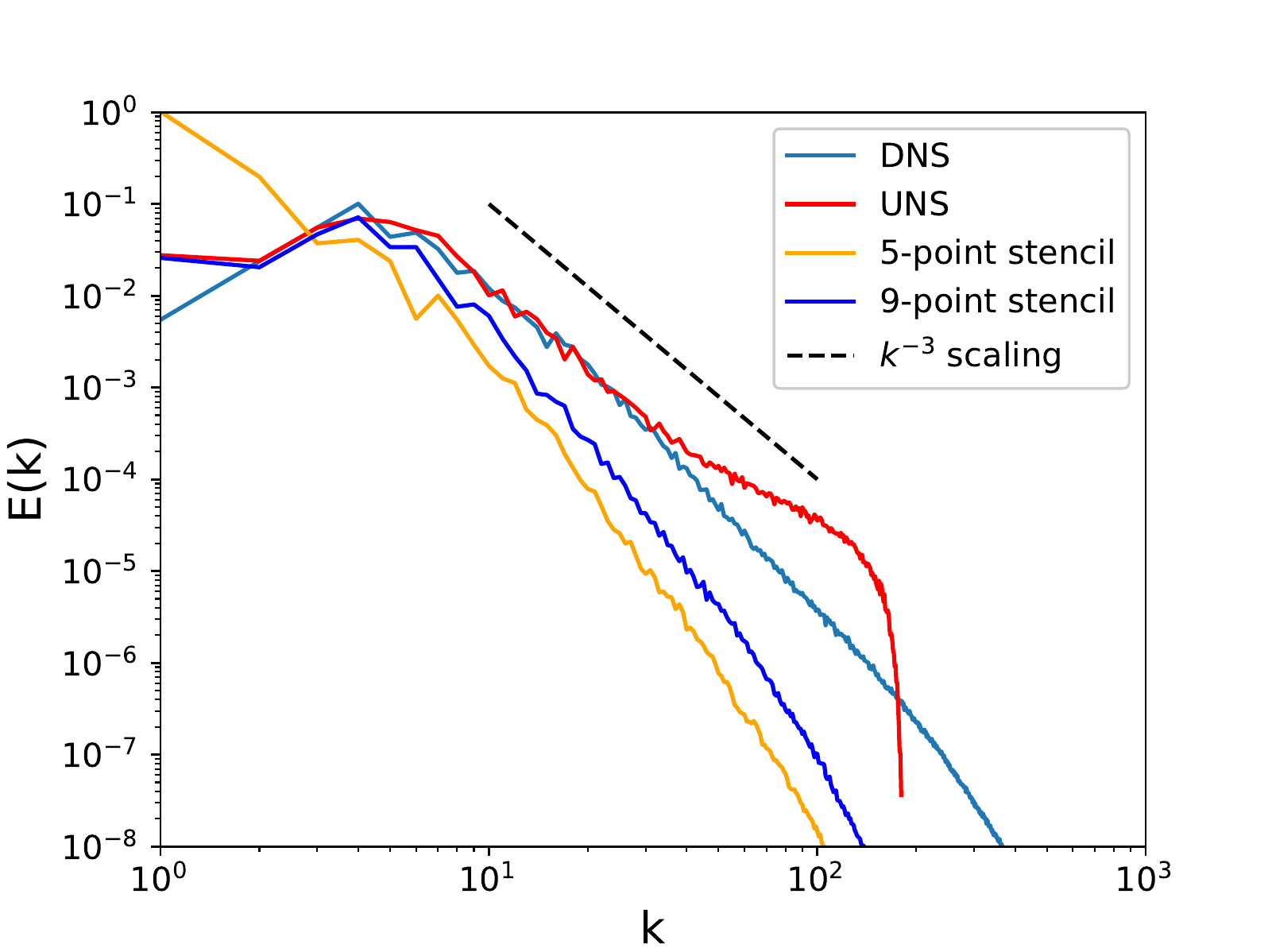}}
}
\caption{A-priori (left) and a-posteriori (right) effect of the stencil size in the 2-layer, 50 neuron framework for a $Re=32000$ simulation. While the 5-point stencil leads to similar a-priori training errors, an a-posteriori deployment at $Re=32000$ reveals its limitations. }
\label{fig:fig14}
\end{figure}

\begin{figure}
\centering
\mbox{
\subfigure[Learning rate]{\includegraphics[width=0.44\textwidth]{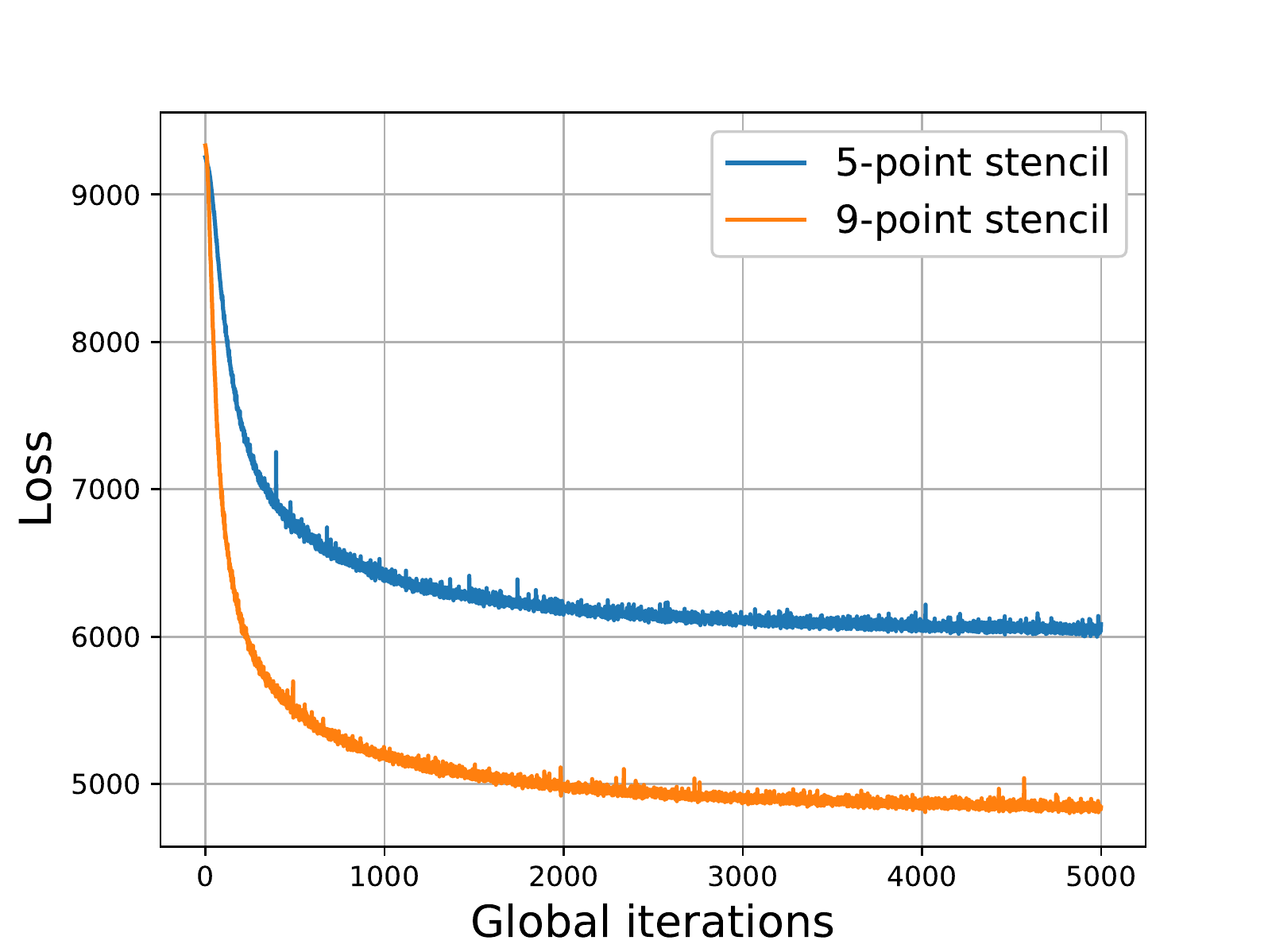}}
\subfigure[A-posteriori deployment]{\includegraphics[width=0.44\textwidth]{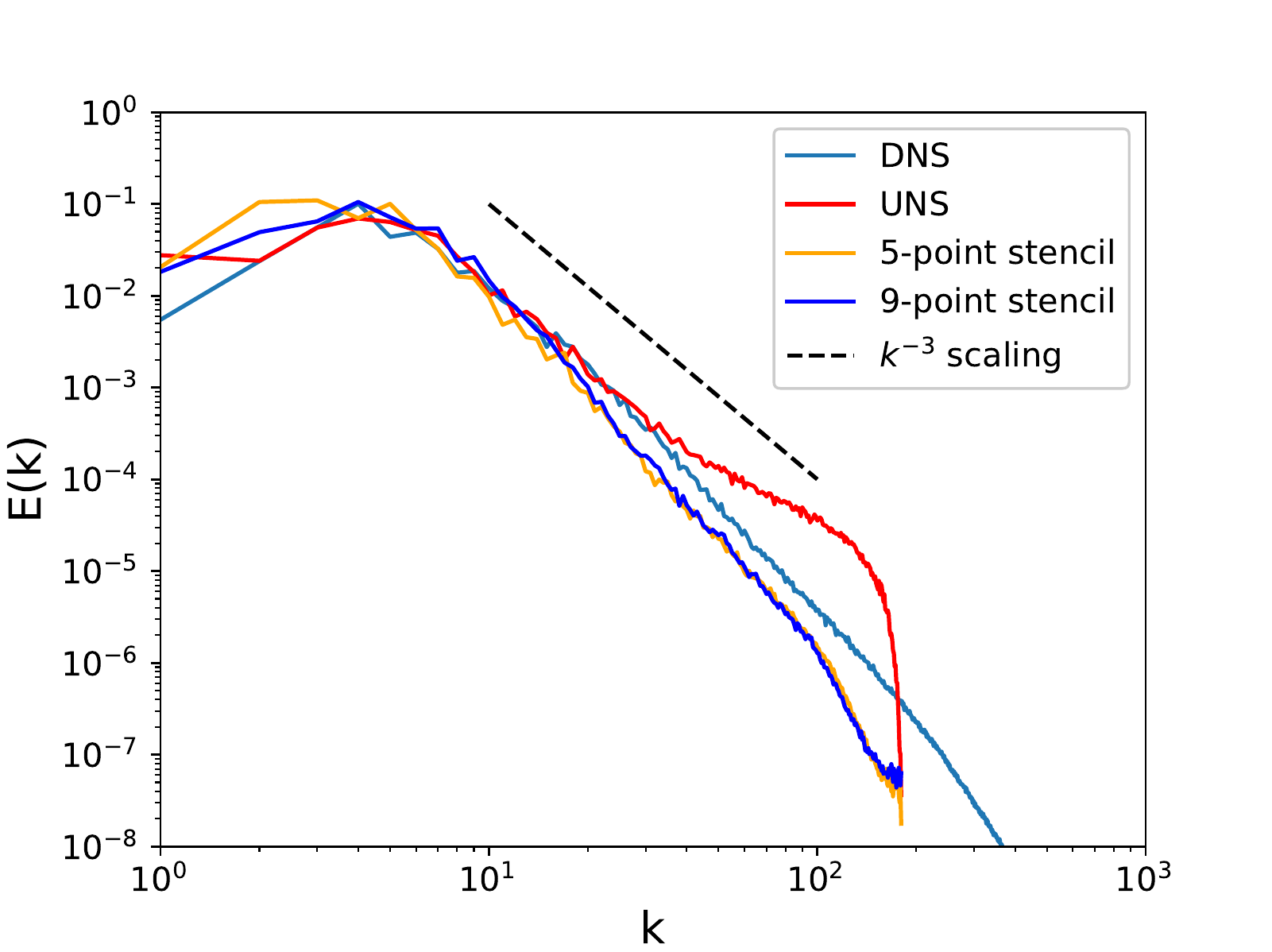}}
}
\caption{A-priori (left) and a-posteriori (right) effect of the stencil size in the 5-layer, 50 neuron framework for a $Re=32000$ simulation. With deeper architectures, the 5 and 9-point stencils show similar statistical performance}
\label{fig:fig15}
\end{figure}

Another comparison is made when the input dimension is substantially reduced by choosing a 5 point stencil (instead of the aforementioned 9 point stencil). In this architecture, vorticity and streamfunction values are chosen only for the $x$ and $y$ directions (i.e., $\bar{\omega}_{i,j},\bar{\omega}_{i+1,j},\bar{\omega}_{i-1,j},\bar{\omega}_{i,j+1},\bar{\omega}_{i,j-1}$ for vorticity and similarly for streamfunction). The input eddy-viscosities given by the Smagorinsky and Leith kernels are also provided to this reduced network architecture. Mathematically, this new map may be expressed as
\begin{align}
\label{eq14}
\begin{gathered}
\mathbb{M} : \{ \bar{\omega}_{i,j}, \bar{\omega}_{i,j+1}, \bar{\omega}_{i,j-1}, \bar{\omega}_{i+1,j}, \bar{\omega}_{i-1,j} \\ \bar{\psi}_{i,j}, \bar{\psi}_{i,j+1}, \bar{\psi}_{i,j-1}, \bar{\psi}_{i+1,j}, \bar{\psi}_{i-1,j}, |\bar{S}|_{i,j}, |\nabla \bar{\omega}|_{i,j}\} \in \mathbb{R}^{12} \rightarrow \{ \tilde{\Pi}_{i,j}\} \in \mathbb{R}^1.
\end{gathered}
\end{align}
Figure \ref{fig:fig14} shows the performance of this setup in training and deployment where it can once again be observed that a-posteriori analysis is imperative for determining a map for the sub-grid terms. While training errors are more or less similar, the reduced stencil fails to capture the nonlinear relationship between the resolved and cut-off scales with consequent results on the statistical fidelity of the lower wavenumbers. We perform a similar study related to this effect of data-locality on a deeper network given by 5 layers and 50 neurons to verify the effect of the deeper architecture on constrained prediction. The results of this training and deployment are shown in Figure \ref{fig:fig15} where it is observed that the increased depth of the ANN leads to a similar performance with a smaller stencil size. This implies that optimal data-locality (in terms of the choice of a stencil) leads to a reduced number of hidden layers. Again, the a-priori mean-squared-error is not indicative of the quality of a-posteriori prediction.

The main take-away from this section thus becomes the fact that optimal architectures and maps for sub-grid predictions require a careful a-priori and a-posteriori study for tractable computational problems (such as the Kraichnan turbulence case) before they may be deployed for representative flows. The effect of realizability constraints and numerical errors often leads to unexpected a-posteriori performance and some form of lightweight deployment must be utilized for confirming model feasibility.

\section{Conclusions}

In this investigation, a purely data-driven approach to closure modelling utilizing artificial neural networks is detailed, implemented and analysed in both a-priori and a-posteriori assessments for decaying two-dimensional turbulence. An extensive hyper-parameter selection strategy is also performed prior to the selection of an optimal network architecture in addition to explanations regarding the choice of input space and truncation for numerical realizability. The motivation behind the search of a model-free closure stems from the fact that most closures utilize empirical or phenomenological relationships to determine closure strength with associated hazards of insufficient or more than adequate dissipation in a-posteriori utilizations. To that end, our proposed framework utilizes an implicit map with inputs as grid-resolved variables and eddy-viscosities to determine a dynamic closure strength. Our optimal map is determined by training an artificial neural network with extremely sub-sampled data obtained from high-fidelity direct numerical simulations of the decaying two-dimensional turbulence test case. Our inputs to the network are given by sampling stencils of vorticity and streamfunction in addition to two kernels utilized in the classical Smagorinsky and Leith models for eddy-viscosity computations. Based on these inputs, the network predicts a temporally and spatially dynamic closure term which is pre-processed for numerical stability before injection into the vorticity equation as a potential source (or sink) of vorticity in the finer scales. Our statistical studies show that the proposed framework is successful in imparting a dynamic dissipation of kinetic energy to the decaying turbulence problem for accurate capture of coherent structures and inertial range fidelity.

In addition, we also come to the conclusion that the effects of prediction truncation (for numerical realizability) and numerical error during forward simulation deployment necessitate the need for a-posteriori analyses when identifying optimal architectures (such as the number of hidden layers and the input spaces). This conclusion has significant implications for the modern era of physics-informed machine learning for fluid dynamics applications where a-priori trained learning is constrained by knowledge from first principles. Our conclusions point toward the need for coupling a-posteriori knowledge during hyper-parameter optimization either passively (as demonstrated in this article) or through the use of custom training objective functions which embed physics in the form of regularization. Our study basically proposes that data-driven spatio-temporally dynamic sub-grid models may be developed for tractable computational cases such as Kraichnan and Kolmogorov turbulence through a combination of a-priori and a-posteriori study before they may be deployed for practical flow problems such as those encountered in engineering or geophysical flows. Studies are underway to extend these concepts to multiple flow classes in pursuit of data-driven closures that may prove to be more universal.

While this article represents the successful application of a proof-of-concept, our expectation is that further robust turbulence closures may be developed on the guidelines presented in this document, with the utilization of more grid-resolved quantities such as flow invariants and physics-informed hyper-parameter optimization. In addition, network-embedded symmetry-considerations are also being explored as a future enhancements for this research. Dataset pre-processing for outlier identification, not utilized in this study, is also a potential avenue for improved a-posteriori performance and more efficient hyper-parameter selection. Our ultimate goal is to determine maps that may implicitly classify closure requirements according to inhomogeneities in a computational domain (through exposure to different flow classes) that may then be ported as predictive tools in multiscale phenomenon with complex initial and boundary conditions. The results in this document indicate a promising first step in that direction.

\section*{Acknowledgement}
The authors are grateful for helpful comments from the referees. We gratefully acknowledge the support of NVIDIA Corporation for supporting our research in this area. The computing for this project was partially performed at the OSU High Performance Computing Center at Oklahoma State University. OS and PV acknowledge the financial support received from the Oklahoma NASA EPSCoR Research Initiation Grant program. AR acknowledges the financial support received from the Norwegian Research Council and the industrial partners of OPWIND: Operational Control for Wind Power Plants (Grant No.: 268044/E20).

\bibliographystyle{jfm}
\bibliography{jfm}

\end{document}